\documentclass[extra]{gji}
\usepackage{multirow} 
\usepackage{timet}

\usepackage{lineno}
\usepackage{float}
\usepackage[fleqn]{amsmath}

\usepackage{mathtools}
\usepackage{graphicx}
\usepackage{url}
\usepackage[utf8x]{inputenc}
\usepackage{booktabs}

\usepackage{placeins}

\let\oldequation\equation
\let\oldendequation\endequation

\renewenvironment{equation}
  {\linenomathNonumbers\oldequation}
  {\oldendequation\endlinenomath}

\usepackage{xcolor}
\definecolor{darkgray}{gray}{0.5}

\definecolor{darkblue}{RGB}{0,0,205}
\usepackage{url}
\usepackage[colorlinks=true]{hyperref}
\hypersetup{
    pdftitle={Evaluating Multi-station Phase Picking Algorithm Phase Neural Operator (PhaseNO) on Local Seismic Networks},
    pdfauthor={Qingkai Kong},
    bookmarks=true,         
    unicode=false,          
    pdftoolbar=true,        
    pdfmenubar=true,        
    pdffitwindow=true,      
    pdfnewwindow=true,      
    linkcolor=darkblue,    
    citecolor=darkblue,        
    filecolor=darkblue,      
    urlcolor=darkblue,           
    bookmarksnumbered=true, 
    pdfdisplaydoctitle=true,
    pdfstartview=FitV,
}

\title{Evaluating Multi-station Phase Picking Algorithm Phase Neural Operator (PhaseNO) on Local Seismic Networks}
\author[Kong et al.]{Qingkai Kong$^1$\thanks{\href{mailto:mail@fwagner.info}{kong11@llnl.gov}}, Avigyan Chatterjee$^1$, Chengping Chai$^2$, Alex Dzubay$^3$, Kayla A. Kroll$^1$, \\ Josh C. Stachnik$^3$, Scott Fertig$^3$, Jeffrey Liefer$^3$ and Paul Friberg$^3$\\
  $^1$ Lawrence Livermore National Laboratory, California, US\\
  $^2$ Oak Ridge National Laboratory, Tennessee, US\\
  $^3$ Instrumental Software Technologies, Inc., New York, US\\
  }

\date{\today}
\pagerange{\pageref{firstpage}--\pageref{lastpage}}
\volume{200}
\pubyear{{\the\year{}}}

\begin{document}

\renewcommand{\thefootnote}{\fnsymbol{footnote}}

\maketitle
\begin{summary}
  Reliable automatic phase picking is important for many seismic applications. With the development of machine learning approaches, many algorithms are proposed, evaluated and applied to different areas. Many of these algorithms are single station based, while recent proposed methods start to combine surrounding stations into consideration in the problem of phase picking. Among these algorithms, the Phase Neural Operator (PhaseNO) shows promising results on regional datasets comparing to existing algorithms. But there are many use cases for the local seismic networks in our community, therefore in this paper we evaluate the performance of PhaseNO on 4 different local datasets and compared the results to PhaseNet and EQTransformer. We used both individual phase picking metrics as well as association metrics to illustrate the performance of PhaseNO. With manually reviewing the newly detected events, we find the PhaseNO model outperformed the single station-based approaches in the local-scale use cases due to its consideration of coherent signals from multiple stations. We also explored PhaseNO’s behaviors when only using one station, as well as gradually increase the number of stations in the seismic network to understand it better. Overall, using the off-the-shelf machine learning based phase pickers, PhaseNO demonstrated its good performance on local-scale seismic networks. 
\end{summary}


\section{Introduction}

Accurate and robust seismic phase picking is a fundamental component of earthquake monitoring with many subsequential seismic applications relying on it. Traditionally, manual phase picking by analysts has served as the foundation for the seismic catalogs. However, it is inherently time-consuming, subjective, and often incomplete, particularly during periods of high seismicity or in regions with sparse seismic networks. Many automated phase picking algorithms have been proposed and widely used in the community such as STA/LTA \citep{Allen1978} and AIC \citep{Takanami1988}. These algorithms provide efficient but noisy first arrival detections. To address these limitations, a range of machine learning-based phase pickers have been developed, that show promising results. With the success of deep learning models in recent years in many fields, they have also significantly improved picking accuracy by automatically extracting features and learning directly from large seismic waveforms \citep{Dokht2019, Johnson2022, Mousavi2020, Ross2018, Zhou2019, Zhu2019, Zhu2022b}. Notable among these, which are widely used in the community, are GPD \citep{Ross2018}, a convolutional neural network structure to pick P and S arrivals, PhaseNet \citep{Zhu2019}, a U-shaped convolutional neural network with skip connections trained to detect P and S arrivals, and EQTransformer \citep{Mousavi2020}, which integrates multi-tasks, i.e, both picking and event detection, in a unified framework. These algorithms have become the go-to phase picking algorithms in many  researchers’ toolbox for the first step to create a seismic catalog. One common feature of these algorithms is that they are all single-station based approaches, which means phases are recognized on one station without considering of nearby stations.
The coherent seismic wave propagation recorded from multiple stations could potentially provide additional information for seismic phase picking, both for improving the detection and accuracy, and potentially generalizing to unseen regions. Therefore, more recently, multi-station approaches have been developed to utilize this feature. S-EqT, the Siamese earthquake transformer (Xiao et al., 2021), combines the feature embeddings using pre-trained EQTransformer from station pairs for phase picking shows improved results on low signal-to-noise ratio seismograms. EdgePhase \citep{Feng2022} uses Graph Neural Network to exchange information relevant to seismic phases that extracted by the encoder of the EQTransformer from nearby stations for phase picking, which also shows promising results. PhaseNO \citep{Sun2023} represents a new generation of phase picker that utilizes the newly developed Fourier Neural Operator \citep{Kovachki2021, Li2020a} and Graph Neural Operator \citep{Li2020b} to leverage the temporal-spatial information from multiple stations in a seismic network with arbitrary geometry to improve pick robustness and accuracy. Initial results have demonstrated strong performance on the 2019 Ridgecrest earthquake sequence from the regional network comparing both to the single station based as well as the multi-station approaches, particularly in noisy and complex waveform environments.
Despite these advancements, a critical gap remains in understanding how such multi-station models generalize across different settings and network configurations. Critically, most of the models have been tested on regional seismic networks that are different from many of the local seismic networks used in private sector. Variations in instrumentation, geology, event characteristics, and noise profiles can significantly affect model performance. 

In this study, we systematically evaluate the performance of PhaseNO across four separate local-scale seismic networks that were deployed to monitor hydraulic fracturing operations. Using a combination of phase picking accuracy, robustness, and association metrics, we compare PhaseNO against both traditional catalogs and leading neural pickers (PhaseNet, EQTransformer) and investigate the implications of its performance in operational and scientific contexts. To better understand PhaseNO, we also conduct sensitivity studies by systematically removing seismic stations to evaluate their relative impact on picking performance.

\begin{figure}
 \centering
 \includegraphics[width=.5\textwidth]{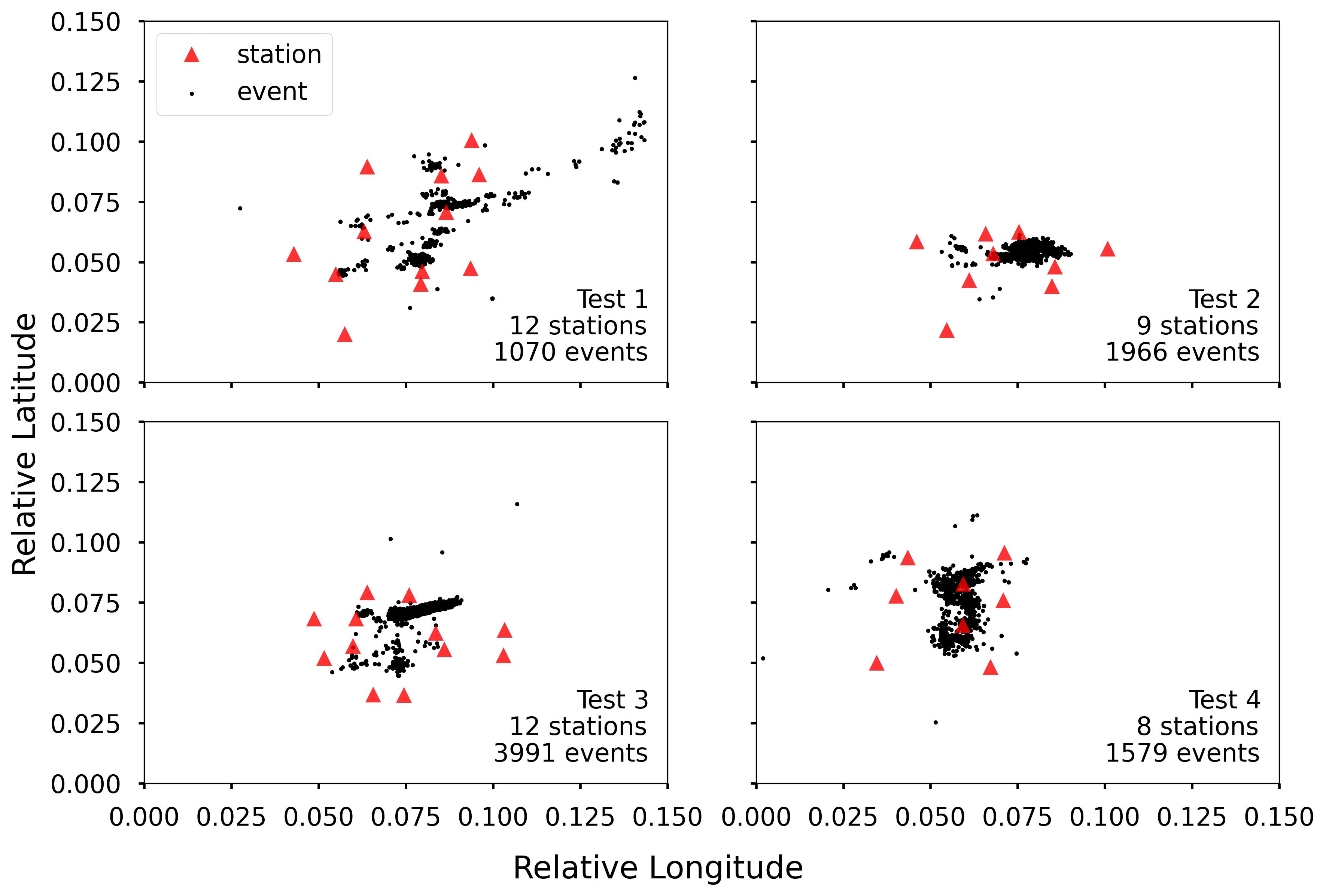}
 \caption{Seismic network configurations as well as the catalog events from 4 different regions. 
 }
 \label{fig:network}
\end{figure}

\section{Data and Networks}

\begin{figure}
 \centering
 \includegraphics[width=.5\textwidth]{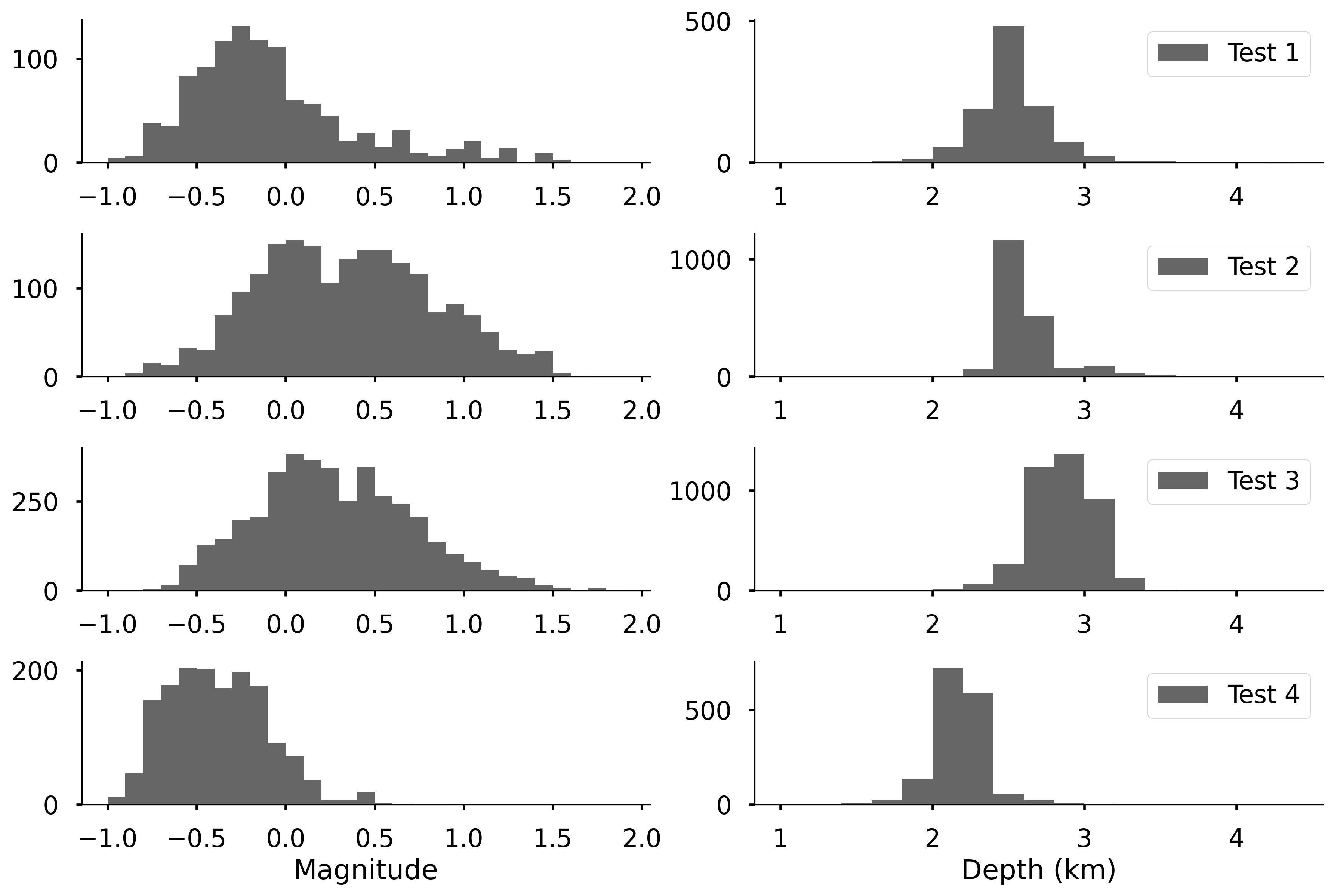}
 \caption{The magnitude, depth and S-P time distribution for the 4 different tests. Each row represents a region.
 }
 \label{fig:data}
\end{figure}

Collaborating with ISTI (Instrumental Software Technologies, Inc.), we obtained 4 different local  datasets for testing the performance of the PhaseNO model. These arrays are composed of 8-12 wideband seismometers installed in shallow post holes with about 1 km spacing. \autoref{fig:network} shows the geometry of the seismic networks as well as the seismic events recorded in these regions. The reference catalogs span 2-4 weeks and were assembled through automatic review, template matching, and manual scoping of waveforms. Automatic review is performed through the automatic processing within Earthworm, template matching using the analyst reviewed events as template to scan and match new detections, while the manual scoping is performed by analyst using a visualization tool to make note of new possible detections. Due to the privacy of the data, we plot only the relative longitude and latitude for both seismicity and stations. The provided waveforms are 120 s windows of 3-component HH channel data with sampling rate 100 Hz. The labels of the manual picked phases for both P and S, as well as the associated information for the corresponding event, are provided. \autoref{fig:data} shows the distributions of magnitude, depth and S-P time of the events in these 4 test regions, most of which are from shallow, low magnitude events that are only recorded by the nearby stations. Due to their smaller magnitudes, it is clear that test 1 and 4 are more challenging than test 2 and 3.

\section{Methodology}
To evaluate the performance of PhaseNO on local seismic networks we compare its performance against two single station based approaches that are widely used: PhaseNet and EQTransformer. While all three models aim to detect and classify seismic phase arrivals from seismic waveforms, they differ significantly in architecture, input representation, training objectives, and inference behavior. In the following, we briefly introduce each algorithm, please refer to the cited original paper for more details. 

\subsection{PhaseNet}
PhaseNet \citep{Zhu2019} uses the U-net structure with residual connections \citep{Ronneberger2015} to detect P and S phases at the sample level. The model takes input from 3-component waveform windows (originally set at 30 seconds) sampled at 100 Hz and normalized channel-wise. Each window is passed through a series of convolutional layers with residual connections and a softmax activation function  to produce a probability distribution over three classes: P, S, and noise. Picks are made by the exceedance of a pre-set threshold in the phase probability output, typically with additional temporal smoothing or thresholding. The model has demonstrated great performance on high-SNR datasets and has been widely used for regional catalog enhancement.

\subsection{EQTransformer}

EQTransformer \citep{Mousavi2020} integrates convolutional layers, bidirectional LSTMs, and attention mechanisms into a hybrid encoder-decoder architecture that jointly performs seismic event detection and phase picking. The model ingests 3-component waveform segments and returns three probabilities over event detection, P arrival and S arrival. The use of recurrent and attention layers enables the model to incorporate long-range temporal dependencies, which improves its sensitivity to emergent arrivals and overlapping events. Its multi-task structure allows for joint training on phase and event labels, often resulting in improved recall for low-magnitude or swarm-like sequences.

We use both the PhaseNet and EQTransformer models from Seisbench \citep{Woollam2022}.

\subsection{PhaseNO}
PhaseNO \citep{Sun2023} departs from single-station paradigms by incorporating multi-station contextual inference via the Fourier Neural Operator (FNO) and Graph Neural Operators (GNO) framework. The model is structured to process full spatiotemporal waveform fields over any network geometries. Instead of making local predictions per trace, PhaseNO learns a mapping from the input waveform fields to a set of phase probability fields, parameterized in the spectral (Fourier) domain. The FNO layers process the temporal information from the waveform fields, while the GNO layer aggregate the information from waveforms at different seismic stations. The FNO layers apply global convolutions in Fourier space, allowing PhaseNO to efficiently capture long-range spatiotemporal correlations across stations. This is particularly advantageous for detecting weak or partially obscured arrivals that are coherent across multiple sensors. 

\subsection{Association algorithm}
We use PyOcto \citep{Munchmeyer2023} to associate all picks from different tests. PyOcto partitions space-time into 4D space-time cells  inspired by the Octotree data structure to associate picks from different stations, only exploring origin regions that are promising, using either homogeneous or 1D velocity models. Within each cell, picks are evaluated for consistency with travel time predictions, and events are generated when cells shrink to a minimum size, with pick refinement and localization applied. PyOcto has been tested on two synthetic datasets (with varying event rates and noise) as well as the 2014 Iquique sequence (dense aftershock sequence), demonstrating on par or even superior results when compared to the existing algorithms, such as GaMMA \citep{Zhu2022a} and REAL \citep{Zhang2019}, but with a substantial improvements in computation time. 

\section{Evaluation Metrics}

To assess the performance of PhaseNO and compare it against baseline models (PhaseNet and EQTransformer), we employ a suite of metrics that evaluate both individual pick quality and downstream impact on seismic event association.

\subsection{Phase Picking Accuracy}

We evaluate phase detection performance using per-pick classification metrics, computed relative to the picks that provided by ISTI, which we treat it as ground truth (GT), the precision (\autoref{eq:precision}), recall (\autoref{eq:recall}) and f1 score (\autoref{eq:f1}) are shown below, where the true positives are the number of PhaseNO detected picks that are matching the GT picks within a temporal tolerance window (e.g., $\pm0.5$ s for both P and S), false positives are the number of detected picks that have no corresponding GT picks, while the false negatives are the number of GT picks that are not matched by the detection. The precision measures the fraction of model picks that correctly match the GT picks, recall indicates the percentage of GT picks that are successfully recovered by the model and f1 score is the harmonic mean of precision and recall. 

\begin{equation}\label{eq:precision}
       precision = \frac{True\ positives}{True\ positives + False\ positives}
\end{equation}

\begin{equation}\label{eq:recall}
       recall = \frac{True\ positives}{True\ positives + False\ negatives}
\end{equation}

\begin{equation}\label{eq:f1}
       f1 = 2\cdot \frac{Precision \cdot Recall}{Precision + Recall}
\end{equation}
\\
For the matched picks, we use the pick residual $\Delta{t}$ to further assess the temporal precision of these picks, as shown in \autoref{eq:residual}. The residuals are shown in the histograms as well as reported with their mean, median, and standard deviation values in the results section. 

\begin{equation}\label{eq:residual}
       \Delta{t} = t_{PhaseNO} - t_{GT}
\end{equation}
\\
We also stratify the true positives, false negatives, and false positives by the estimated signal noise ratio (SNR) to evaluate their trade-offs across different quality bins. 

\subsection{Association and Event-Level Metrics}

When picks are passed to a phase associator (e.g., PyOcto), we evaluate how well they associate compared to the reference catalogs based on the GT picks. We again used the precision \eqref{eq:precision}, recall \eqref{eq:recall} and f1 score \eqref{eq:f1} for general evaluation, with the true positives representing the number of PhaseNO detected events that match the GT events in terms of the origin time within a temporal tolerance window (e.g., $\pm1$, $\pm3$, $\pm5$ s for the origin time are evaluated), the false positives representing the number of detected events that have no matched events in the GT catalogs, and the false negatives representing the number of GT events that have no matched detected events. 

Metrics such as origin time, epicentral distance, depth, and magnitude errors against the GT catalog events are also provided for further evaluations. 

\subsection{False Positive and Novel Detection Validation}

Because PhaseNO often generates more picks and thus resulted in more detected events than the GT catalog, we use two additional validation tests to characterize the unmatched events: first, the automatic cluster-based validation test, where detected picks are grouped using spatial-temporal clustering (e.g., via PyOcto). For the events that have more picks (multiple stations detected), it is more likely to be real, we use a threshold of 10 triggers within the event as a threshold for determining the potential real event within the new detections. Second, manual review sampling validation test, where we randomly sample a subset of 100 newly detected events from each test region, which then are reviewed by the authors to estimate the proportion of valid but uncataloged events.

\section{Results}

\subsection{Phase Picking Metrics}

\begin{figure}
 \centering
 \includegraphics[width=.5\textwidth]{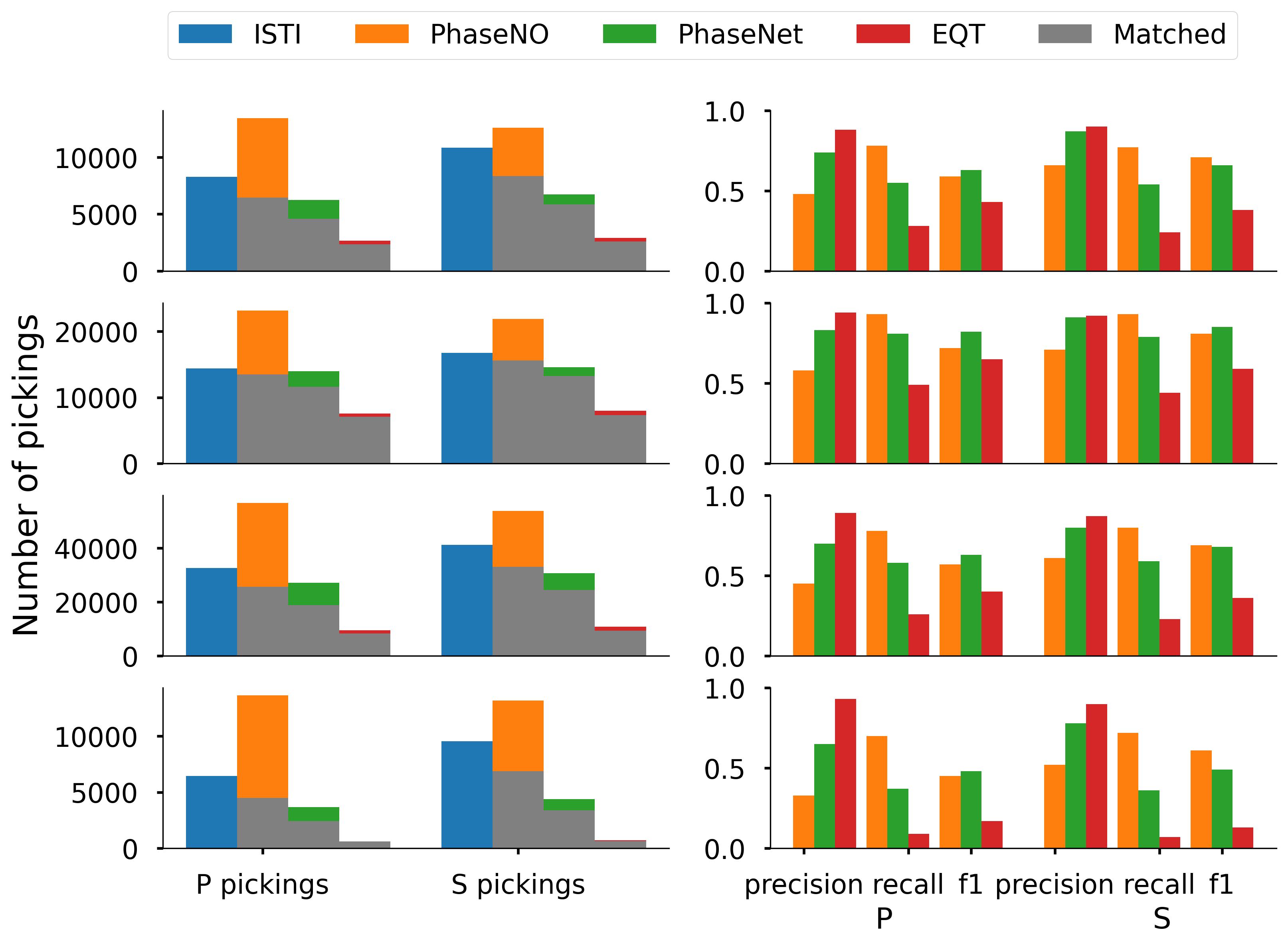}
 \caption{Phase picking metrics, the left panels show the histogram comparisons for different algorithms, the greyed histograms are the matched phases for each algorithm. The right panels show the precision, recall and f1 metrics for both P and S pickings. The rows from top to bottom are test 1 to test 4. The detailed numbers are listed in \autoref{tab:A1}. 
 }
 \label{fig:pickingMetrics}
\end{figure}

Metrics on individual picks from the different algorithms can provide us the first direct view on how well each algorithm identifies both P and S waves. \autoref{fig:pickingMetrics} shows the number of pickings from each algorithm compared to the GT picking list, as well as the precision, recall and f1 scores. From these metrics, we can see that PhaseNO detects considerably more arrivals compared to PhaseNet and EQT, and thus matched more reference picks. This also reflects in the recall scores, where PhaseNO has the highest scores, followed by PhaseNet and EQT. However, PhaseNet and EQT have higher precision scores, meaning the picks they detect are more likely to match with the reference list. From the f1 score point of view, the PhaseNet model has the highest performance on P pickings, closely followed by PhaseNO, and then EQT. The S-picking f1 scores are ranked in the order of PhaseNO, PhaseNet and EQT. Overall, from these metrics, it seems PhaseNO and PhaseNet are similar in performance, but highlight differently on precision and recall while EQT is less competent. 

\begin{figure}
 \centering
 \includegraphics[width=.5\textwidth]{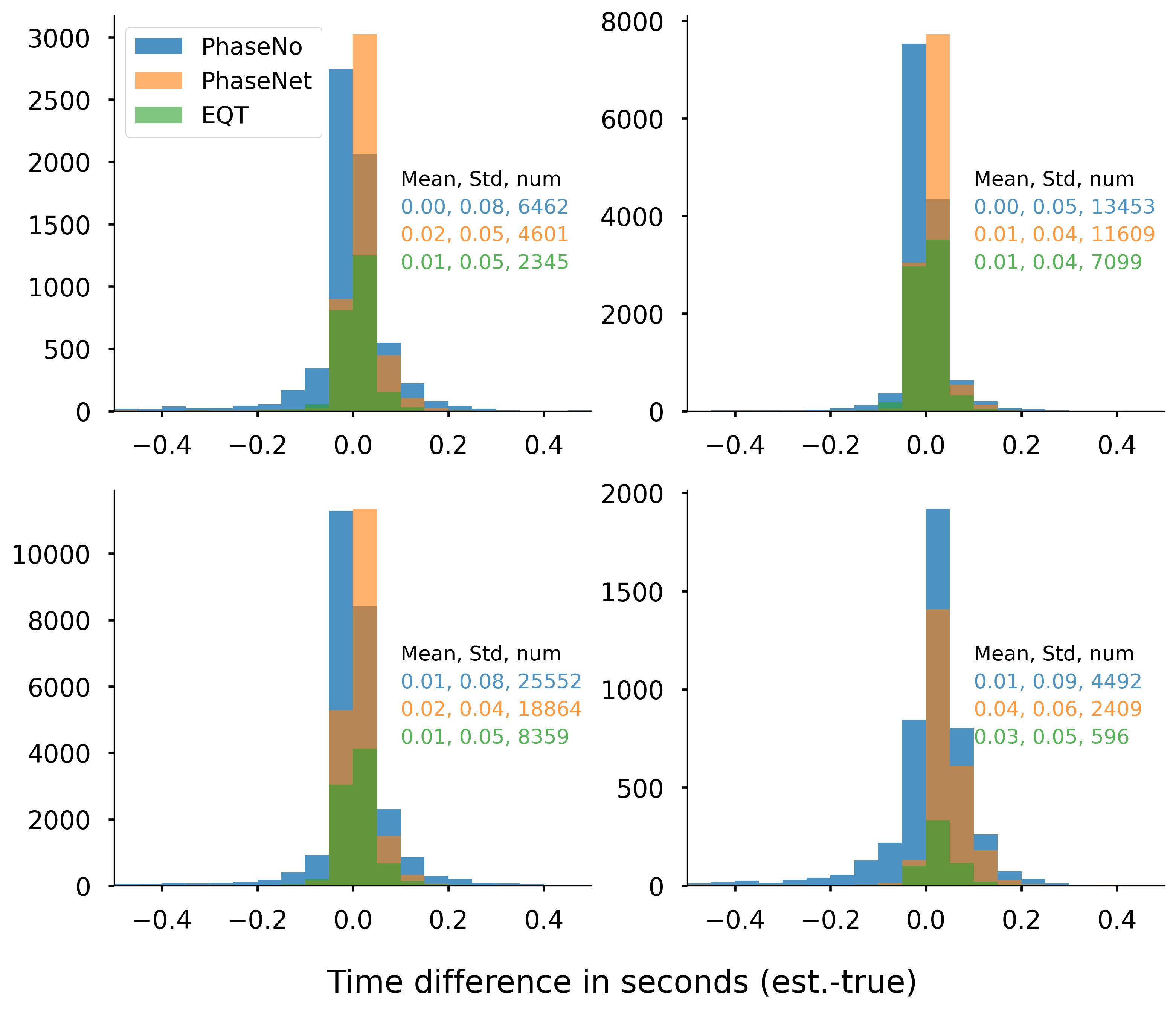}
 \caption{P wave picking time differences comparing to manual pickings. Mean, standard deviation and number of phases are showing in each panel. Top row panels are test 1 and test 2 from left to right, while the bottom row has test 3 and test 4. Similar figure for S wave picking differences is shown in \autoref{fig:A1}. Detailed numbers are listed in \autoref{tab:A2}. 
 }
 \label{fig:residual}
\end{figure}

\autoref{fig:residual} examines the P pick time differences comparing to the ISTI’s provided picks. We use the threshold 1s to match the picks. All three algorithms have similar mean values around 0, while the PhaseNO has larger standard deviation. This is likely due to PhaseNO generating more phase picks, and therefore a wider spread. However, PhaseNO also has more picks with a time difference nearzero compared to the other two algorithms, which puts PhaseNO at an advantage. 

\begin{figure*}
 \centering
 \includegraphics[width=0.8\textwidth]{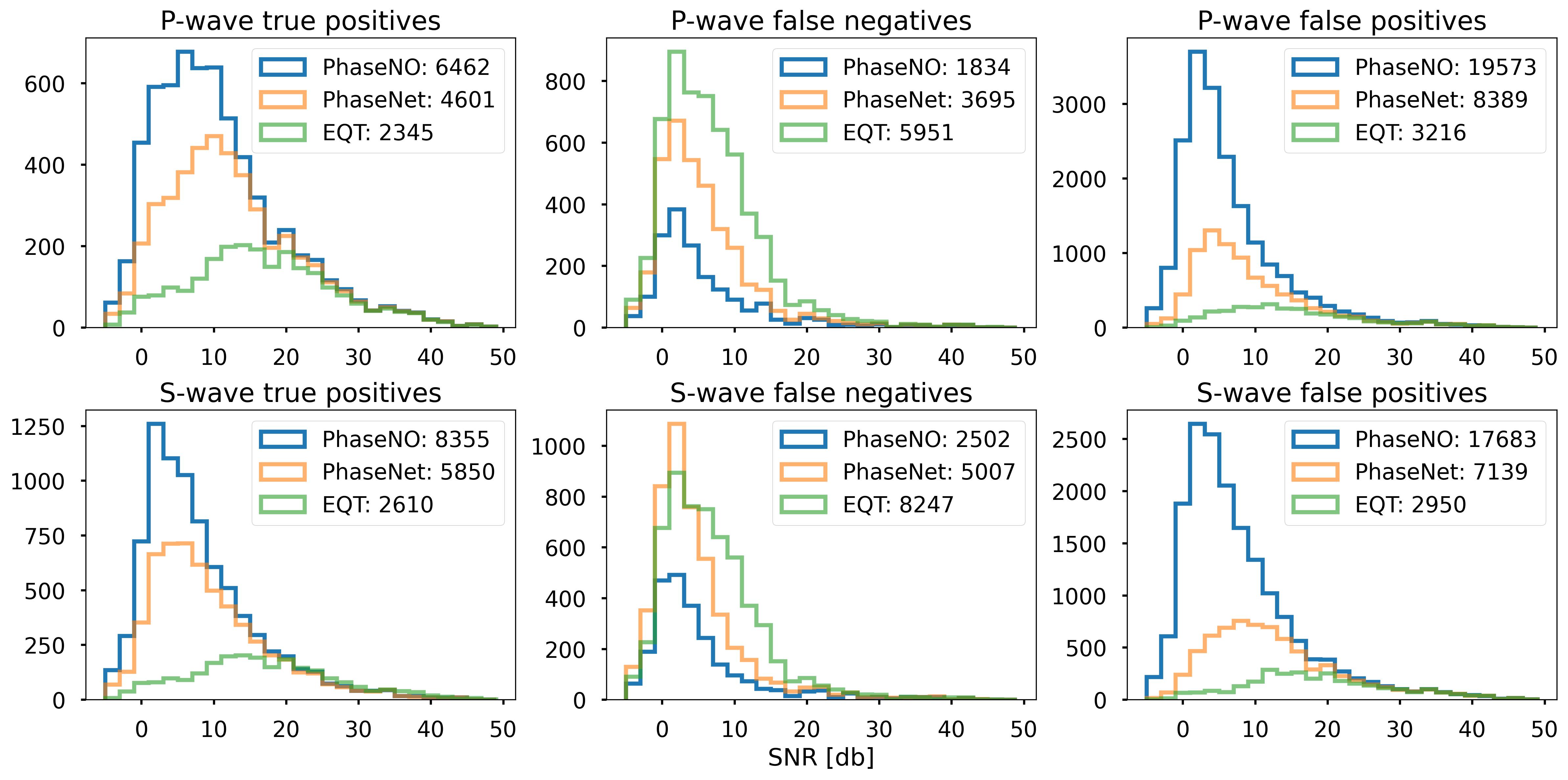}
 \caption{Test 1 phase true positives, false negatives, and false positives distributions for both P and S phases against signal noise ratio. For test 2-4, please refer to \autoref{fig:A2} - \autoref{fig:A4}.
 }
 \label{fig:snr}
\end{figure*}

In order to better understand of the performance of the different algorithms under various signal-to-noise ratios (SNR), we stratify the true positives, false negatives, and false positives by the estimated SNR (we are using 3s windows before the P picks as noise windows and 0.5s after the pickings as the signal windows). \autoref{fig:snr} plots the results. For the true positives, the three models perform similarly when the arrivals have high SNR. Differences arise from the low SNR arrivals. While EQT starts to lose the sensitivity around a SNR of 15 dB and PhaseNet around 10 dB,  PhaseNO retrieves far more picks with SNR less than 10 dB. This shows PhaseNO’s ability to extract useful picks from low SNR signals. For the false negatives (where the models miss the detection of the true pickings), PhaseNO performs the best, with the majority of its missing picks being low-amplitude arrivals with SNR less than 5 dB. It rarely misses  picks with SNR larger than 15 dB. The results of the false positive test are different than the previous two. Since PhaseNO detects much more picks, it has a much larger false positive rate than the other two algorithms, though most of the false positive picks are low SNR. We will revisit the false positives in the next section through the view of association, where we find that a big chunk of the newly detected picks are from the real earthquakes. Again, at high SNR above 20 dB, the three algorithms perform similarly.

\subsection{Association metrics (vs. GT Catalogs)}

\begin{figure}
 \centering
 \includegraphics[width=.5\textwidth]{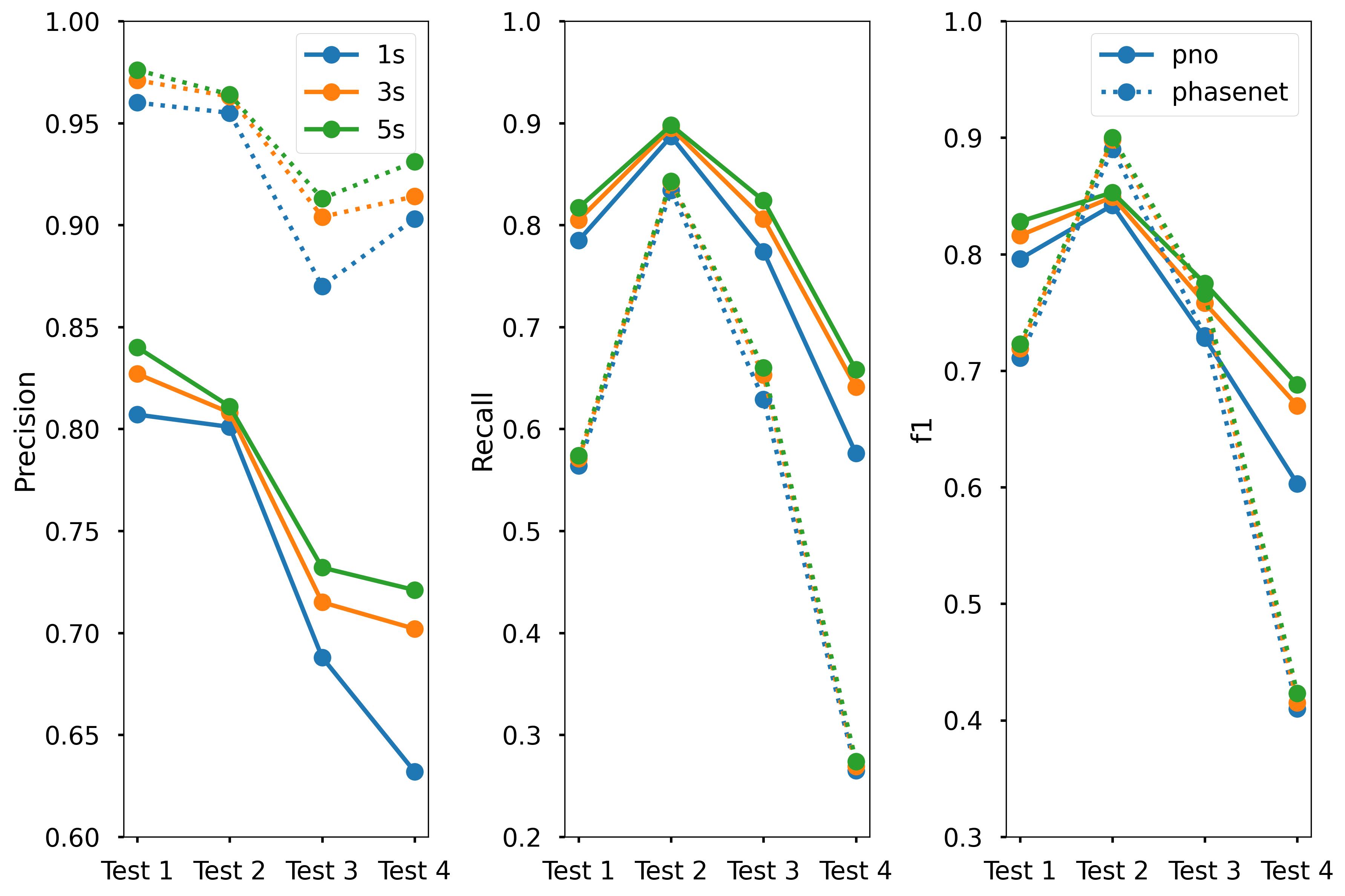}
 \caption{Precision, recall and f1 for associated events for the 4 different tests. Different colors are indicating the matching threshold using origin time (1, 3, 5s), the solid lines are from PhaseNO results, while the dotted lines are from PhaseNet.
 }
 \label{fig:prf1_events}
\end{figure}

Individual phase pickings from different algorithms can be associated based on their spatial and temporal natures to form seismic event. Using PyOcto, we associate the picks from the 4 test regions into seismic events and compare them with the GT catalogs provided by ISTI. \autoref{fig:prf1_events} shows the precision, recall and f1 score for PhaseNO and PhaseNet with different matching thresholds, i.e. 1, 3, 5s. Similar to the phase metrics, the PhaseNO model scores well on recall and poorly on precision compared to PhaseNet. That said,  PhaseNO does produce f1 scores higher than PhaseNet in all of the tests except for  test 2. Therefore, from this metrics, the PhaseNO model works better for the local events. 

Different matching thresholds do not affect the metrics too much for test 2, but  test 1 and 4 have more noticeable differences for PhaseNO. \autoref{tab:A3} documents the specific statistics for the association results with a matching threshold of 1s. For PhaseNO, overall, about 75\% of the picks are associated to form events except for test 4 with only 63\% of the total picks.

\begin{figure}
 \centering
 \includegraphics[width=.5\textwidth]{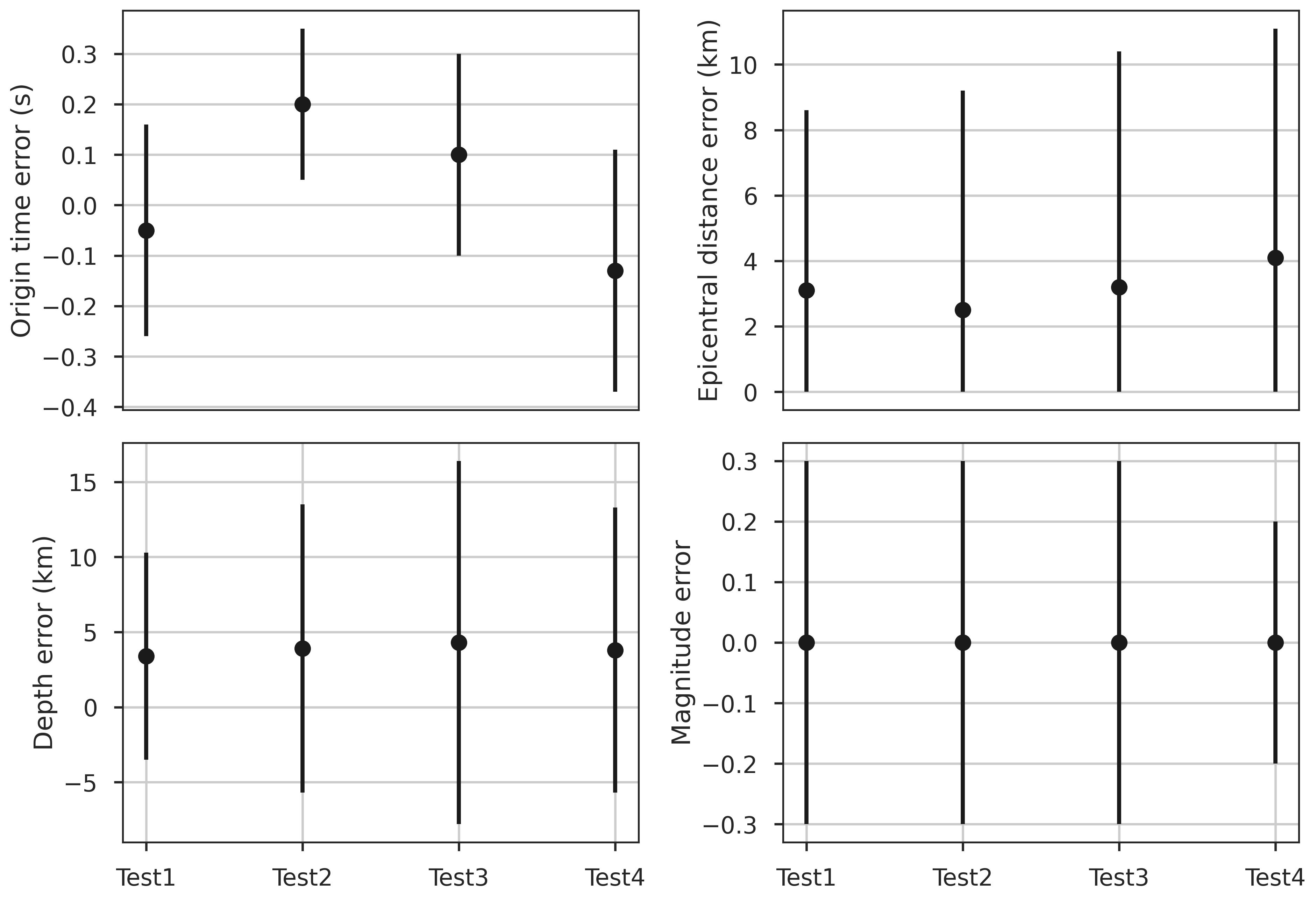}
 \caption{Matched events residuals from PhaseNO model against the provided catalogs (with matching threshold 1s). The dots indicate the mean values while the vertical black lines are standard deviations. For detailed histograms, please refer to \autoref{fig:A5} - \autoref{fig:A8}.
 }
 \label{fig:eventMetrics}
\end{figure}

Because different matching thresholds don’t change the results too much, in the following, to derive the metrics for the matched events comparing to the catalog, we only show the results from the matching threshold 1s. \autoref{fig:eventMetrics} shows the errors of origin time, epicentral distance, depth and magnitude for the matched events for all 4 different regions. The mean errors for different parameters are generally small, with some of them have relatively large standard deviation.

\begin{figure*}
 \centering
 \includegraphics[width=0.8\textwidth]{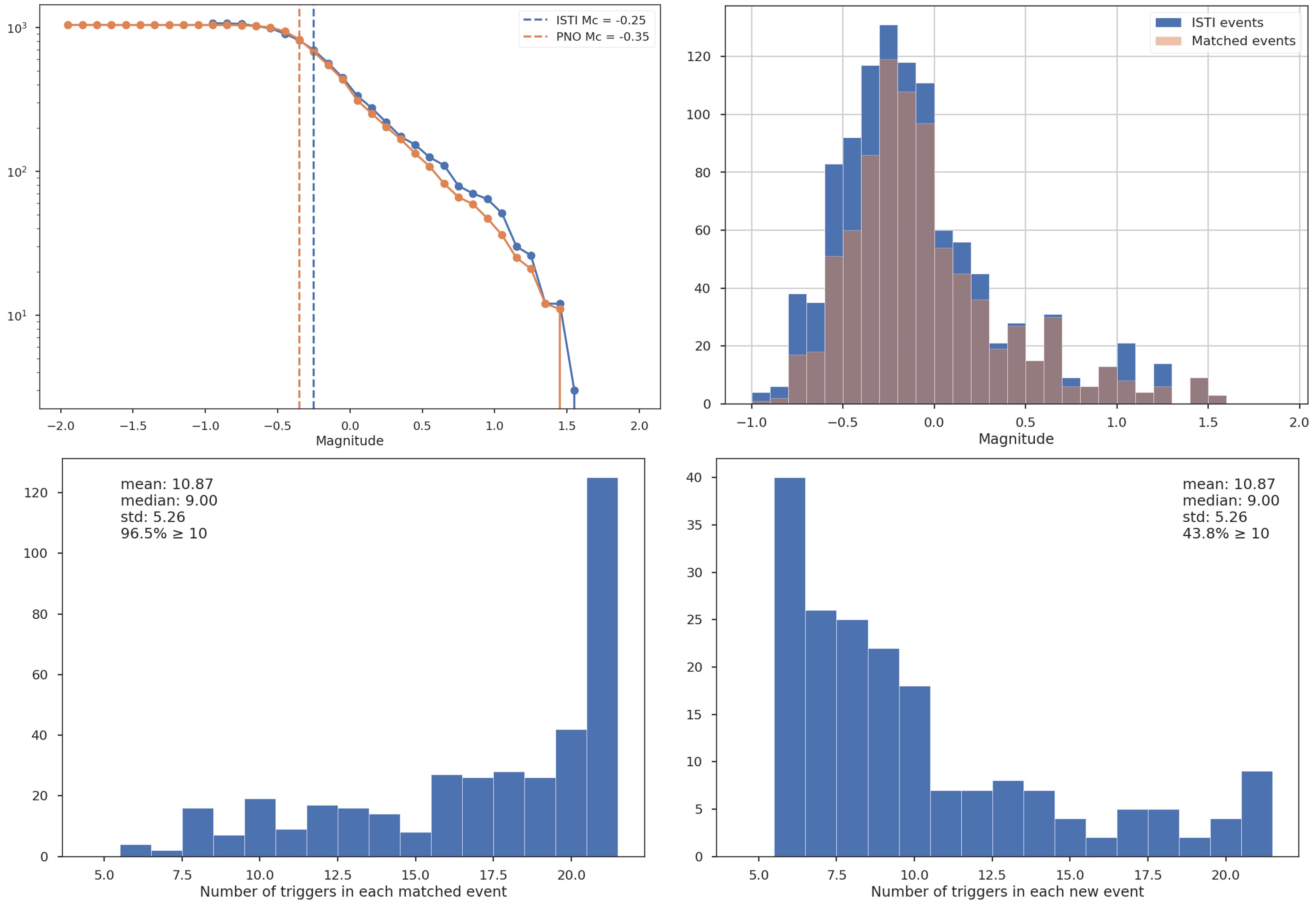}
 \caption{More metrics for the associated events for test 1. Top left panel shows the magnitude GR relationship and magnitude of completeness for both PNO and ISTI catalogs. The top right panel shows the magnitude distribution for ISTI catalog as well as for the matched events from PNO. The bottom two panels show the trigger distributions for each matched events (96.5\% events have more than 10 triggers) and the new detected events (43.8\% events have more than 10 triggers). For the rest of the test regions, please refer to \autoref{fig:A9} - \autoref{fig:A11}.
 }
 \label{fig:matchedErrors}
\end{figure*}

In \autoref{fig:matchedErrors}, we also show the magnitude Gutenberg-Richter (GR) relationship with the magnitude of completeness for test 1. Together with the figures from \autoref{fig:A9}-\autoref{fig:A11}, we can see that the associated events from PhaseNO decrease the magnitude of completeness for 3 out of 4 regions. From the histograms comparing the matched events with the catalog events against magnitude, we can see that the majority of the missing events are small magnitude events from each test dataset. The two bottom panels from \autoref{fig:matchedErrors} show the number of picks distribution for all the matched events as well as the newly detected events. We find that the majority of the matched events contain more than 10 picks. This number is a good indicator for determine whether an event is real or not. For example, in test 1, more than 96\% of the matched events have more than 10 picks. Using this threshold, it follows that nearly 44\% of the newly detected events could potentially be real events that are missing from the GT catalog. 

\subsection{False Positive and Novel Detection Analysis}

\begin{figure}
 \centering
 \includegraphics[width=.4\textwidth]{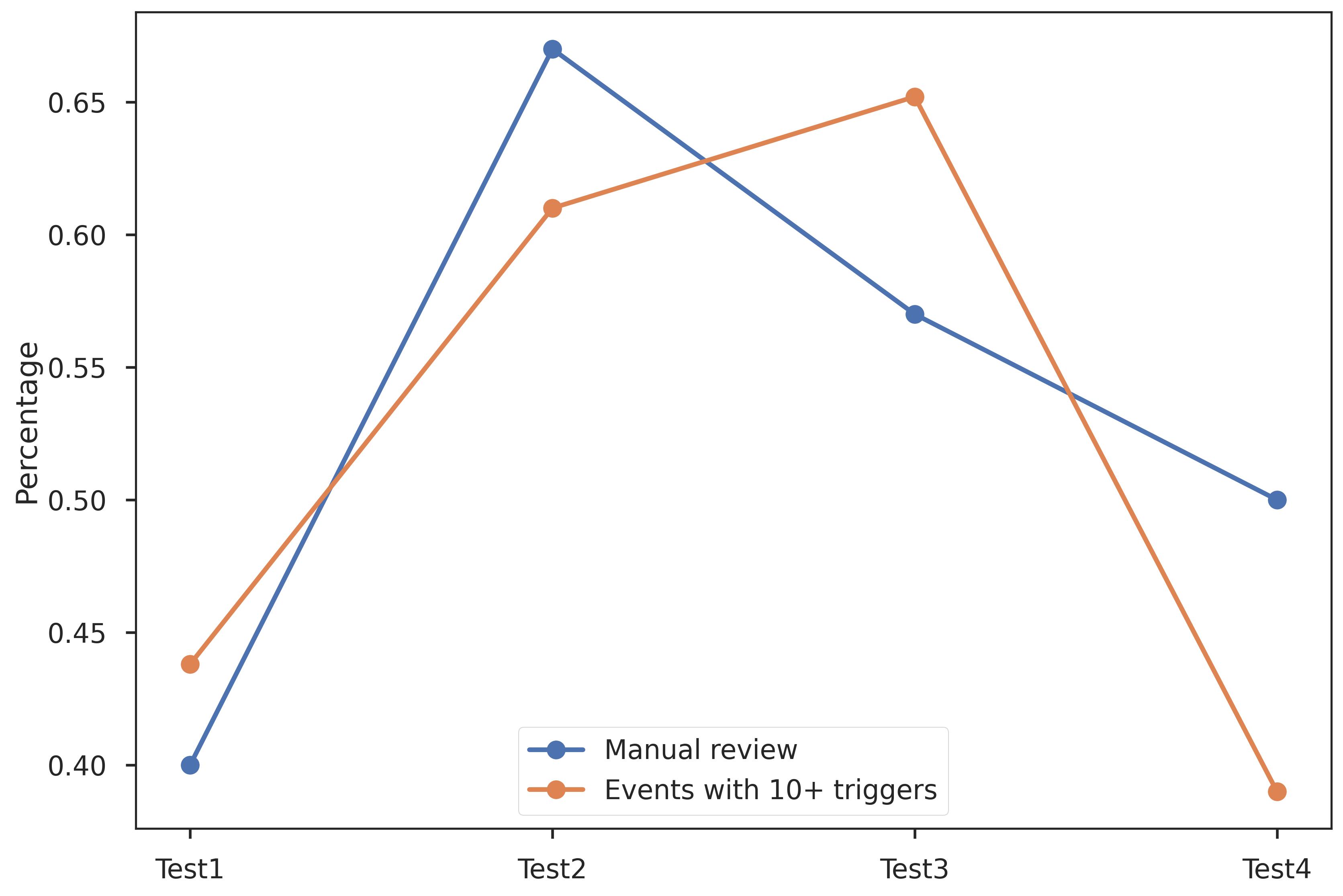}
 \caption{Percentage of potential real events from the unmatched PhaseNO detected events based on manual checking (100 events per test) and using the number of associated triggers as indicator (10+ triggers). 
 }
 \label{fig:newEventMetrics}
\end{figure}

\begin{figure}
 \centering
 \includegraphics[width=.4\textwidth]{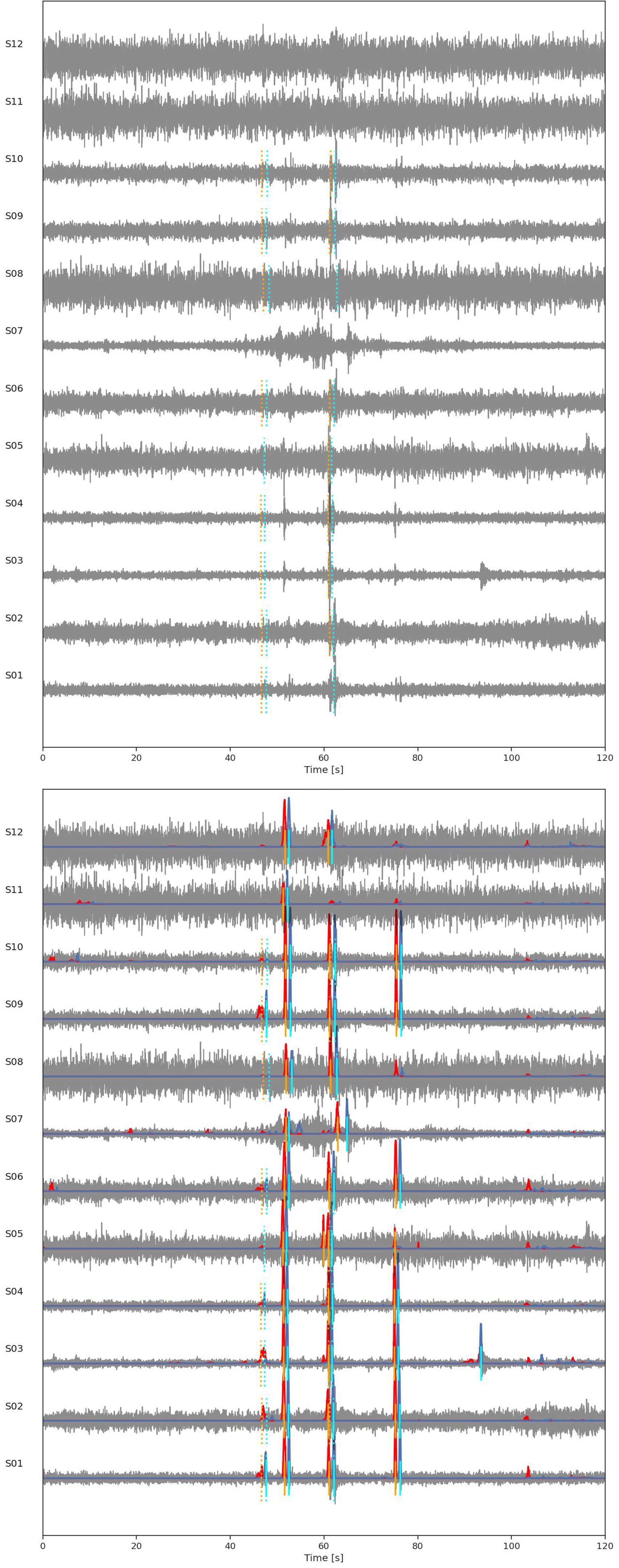}
 \caption{Example of newly detected events. Top panel shows the waveforms from all the stations in test 1, the orange and cyan dotted lines are the ground truth P and S phases. The bottom panel shows the estimated phases from PNO, with solid orange and cyan lines are P and S phases. The red and blue curves are the PNO output probabilities for P and S pickings. The picking of the phases by PhaseNO are showing in orange and cyan solid lines for P and S. The two new detections are around 52s and 76s. 
 }
 \label{fig:waveformExample}
\end{figure}

Since PNO detected new events that are not found in the reference catalog, we treat them as false positives, which lowers its precision as shown in \autoref{fig:prf1_events}. To better understand the portion of true events within these newly formed events, we randomly sampled 100 events from each test region and manually reviewed these events. We plot the results as the blue dots in \autoref{fig:newEventMetrics}. Test 2 and 3 show the highest percentage of real events from these new detections and even test 1 and 4 show about 40\% of the new detections as real events. This additional analysis indicates that the false positive metrics for both phase and association are actually biased. Should these new events have been found within the GT catalog, we’d expect a higher precision and f1 score for the PhaseNO model, which will place PhaseNO to an even higher position compared to other models. In \autoref{fig:newEventMetrics}, we also plot the percentage of the potential real events within each region using the metric of 10+ picks in each newly detected event (the orange dots). We can see they are not so far from the manual checked results, indicating that this value could be used for rough estimation of the real events within the new detections. \autoref{fig:waveformExample} gives an example of the newly detected events. In the GT catalog, there are only two events with the associated phases shown in the top panel. However, PhaseNO detects two more events around 52s and 76s, where we can clearly see coherent signals across multiple stations with relatively low SNR. Many of the newly detected cases are like this example, with some other events counted as new detections due to origin time differences exceed the matching threshold (in this case, we use 1s as threshold). 

\section{Conclusion and Discussion}

In this paper, we evaluated the multi-station approach PhaseNO on local-scale datasets and compaed it to existing phase picking algorithms, such as PhaseNet and EQTransformer. From individual phase metrics to the association metrics, PhaseNO \ demonstrates superior performance when compared to the single station based approaches i.e., it detects more events, matched more events, works better for the low SNR events and so on. 

\begin{figure}
 \centering
 \includegraphics[width=.4\textwidth]{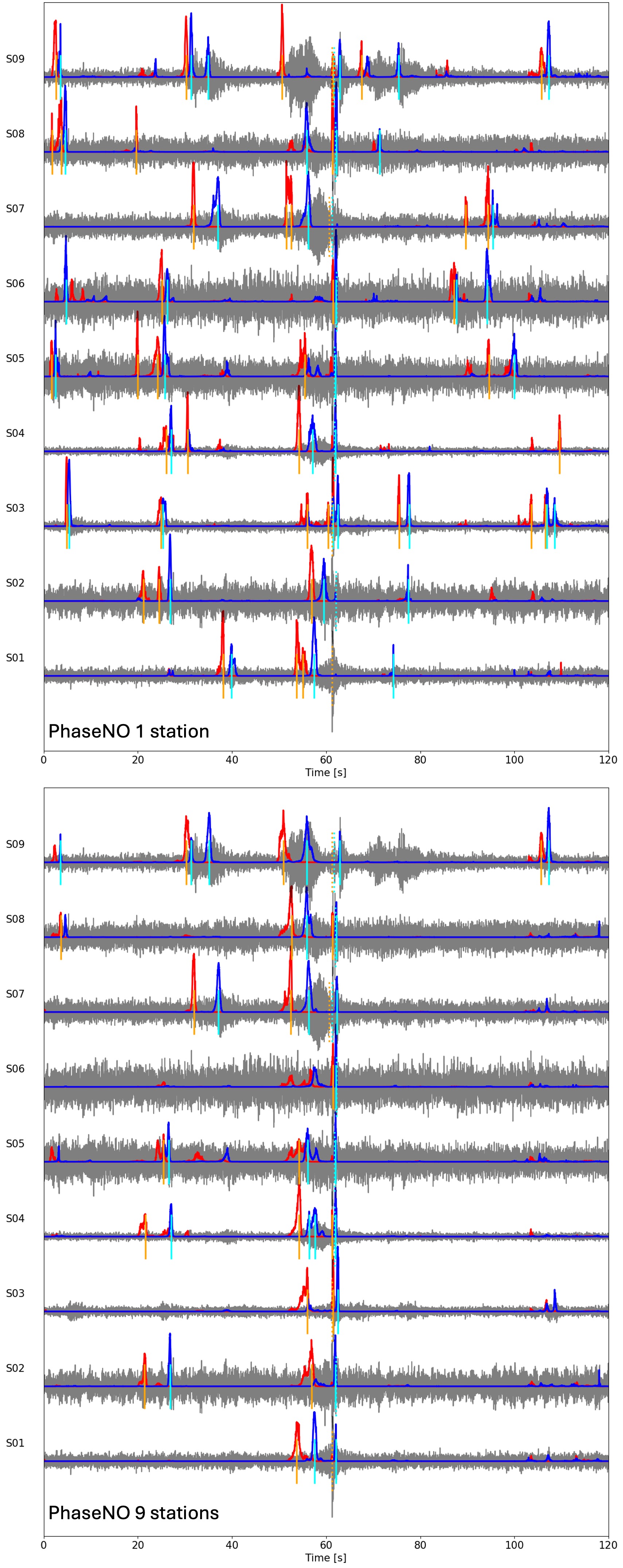}
 \caption{Examples of PhaseNO using 1 and 9 stations. Top panel shows the waveforms and P and S picks when PhaseNO only have one station for phase picking. The bottom panel shows the results for PhaseNO has 9 stations as input. The orange and cyan dotted lines are the ground truth P and S phases from ISTI’s catalog and solid lines are the determined phases from PhaseNO models, while the red and blue curves are the model output probabilities for P and S pickings. 
 }
 \label{fig:expPNO1}
\end{figure}

\begin{figure}
 \centering
 \includegraphics[width=.5\textwidth]{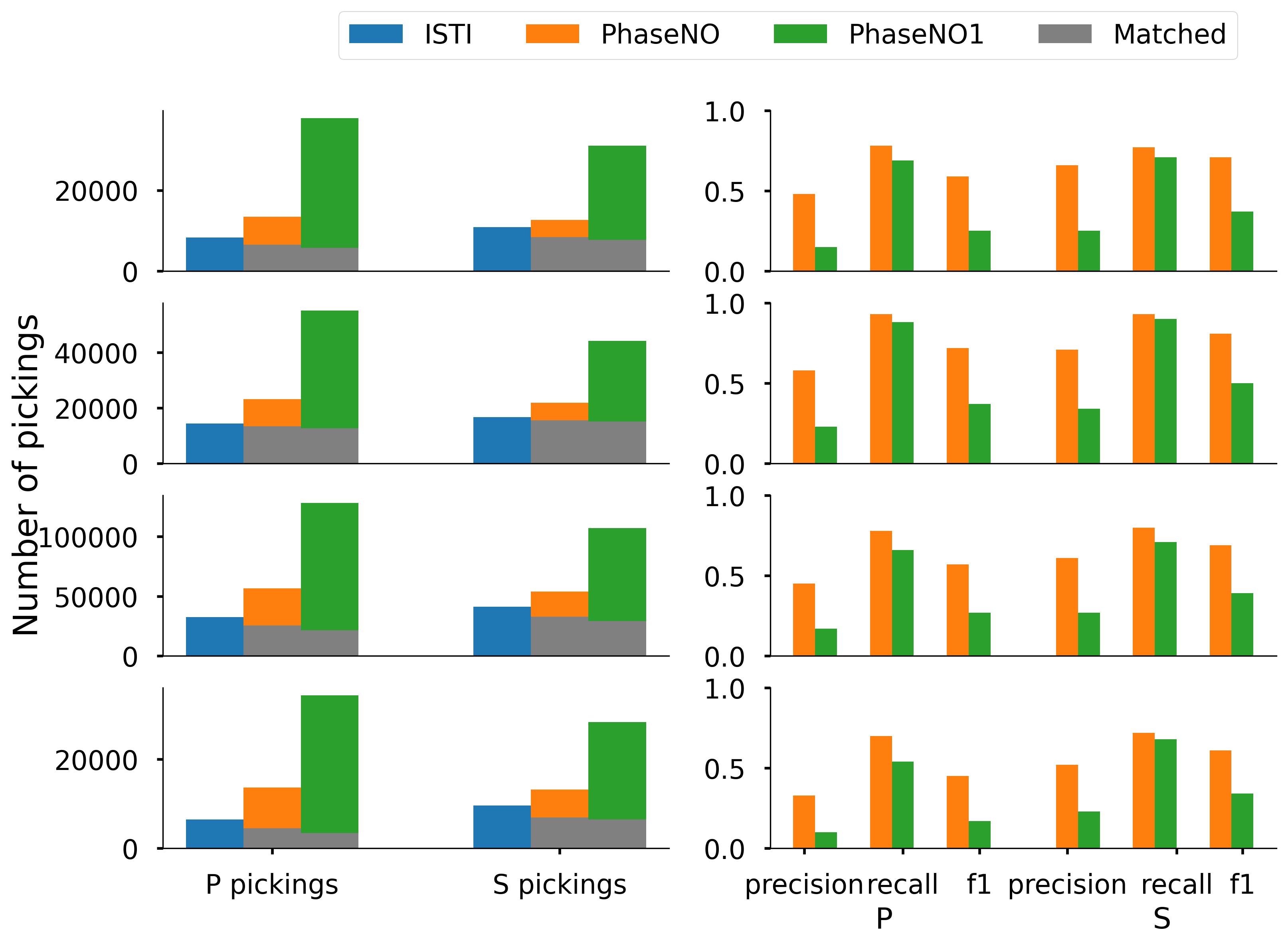}
 \caption{Phase picking metrics comparison between PhaseNO1 and PhaseNO, the left panels show the histograms of the P and S picks, the greyed histograms are the matched phases for each algorithm. The right panels show the precision, recall and f1 metrics for both P and S pickings. The rows from top to bottom are test 1 to test 4. The detailed numbers can be found in \autoref{tab:A4}. 
 }
 \label{fig:PNO1Metrics}
\end{figure}

\begin{figure}
 \centering
 \includegraphics[width=.5\textwidth]{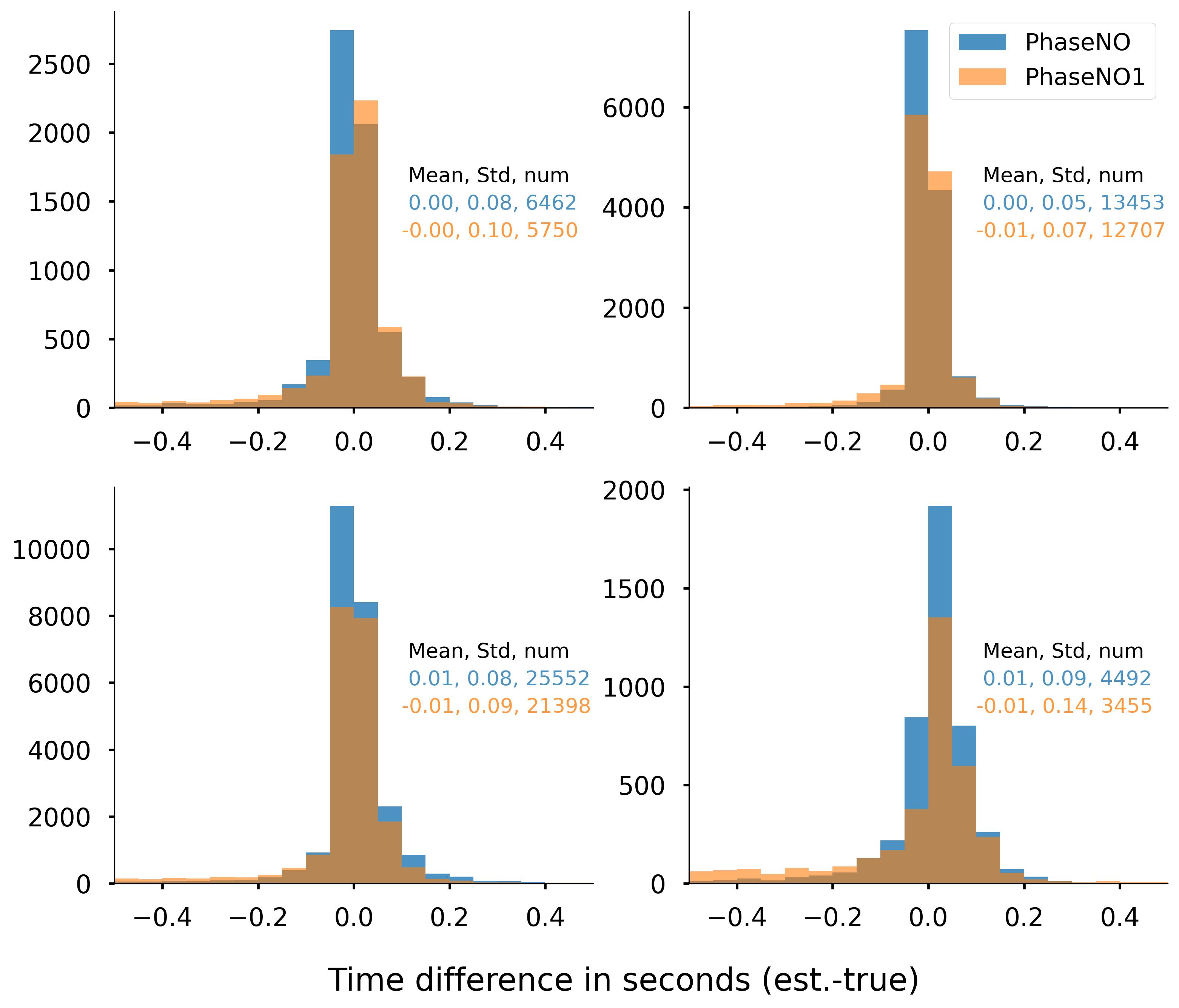}
 \caption{P wave picking time differences comparing to manual pickings between PhaseNO and PhaseNO1. Mean, standard deviation and number of phases are showing in each panel. Top row panels are test 1 and test 2 from left to right, while the bottom row has test 3 and test 4. 
 }
 \label{fig:PNO1Residual}
\end{figure}

To better understand the FNO’s working mechanism, we purposely test PhaseNO model as a single station approach (we call this model PhaseNO1), that is, the input only contains one station waveform, and process all the stations in this way. \autoref{fig:expPNO1} gives a comparison between PhaseNO with only one station as input and that with 9 stations as input. We can see the PhaseNO with one station as input tends to pick more phases at places where there is  little signal, while PhaseNO with multiple stations as input will skip these phases when there are no phases appear on nearby stations. This is expected, as more station waveforms provide more constraints to pick the phases, so the localized energy group will not cause confusions of the model. \autoref{fig:PNO1Metrics} provides the metrics between PhaseNO and PhaseNO1. Even though the PhaseNO1 model detects more phases (more than 2x), the matched P and S picks are slightly decreased comparing to the PhaseNO model results, thus resulting in reduced recall values. After checking the reduced number of matched phases, we find that the reason is due to PhaseNO1 picking phases at slightly different timestamps, and therfore  not satisfying the matching threshold of 0.5s. This indicates that, with multi-station waveforms and PhaseNO’s processing capability, PNO is not only improving its detection rates by seeing coherent signals across multiple stations, but also the accuracy of the pickings. This observation is confirmed by \autoref{fig:PNO1Residual}, where the time residuals of the matched phases are less spread for the PhaseNO model. 

\begin{figure}
 \centering
 \includegraphics[width=.4\textwidth]{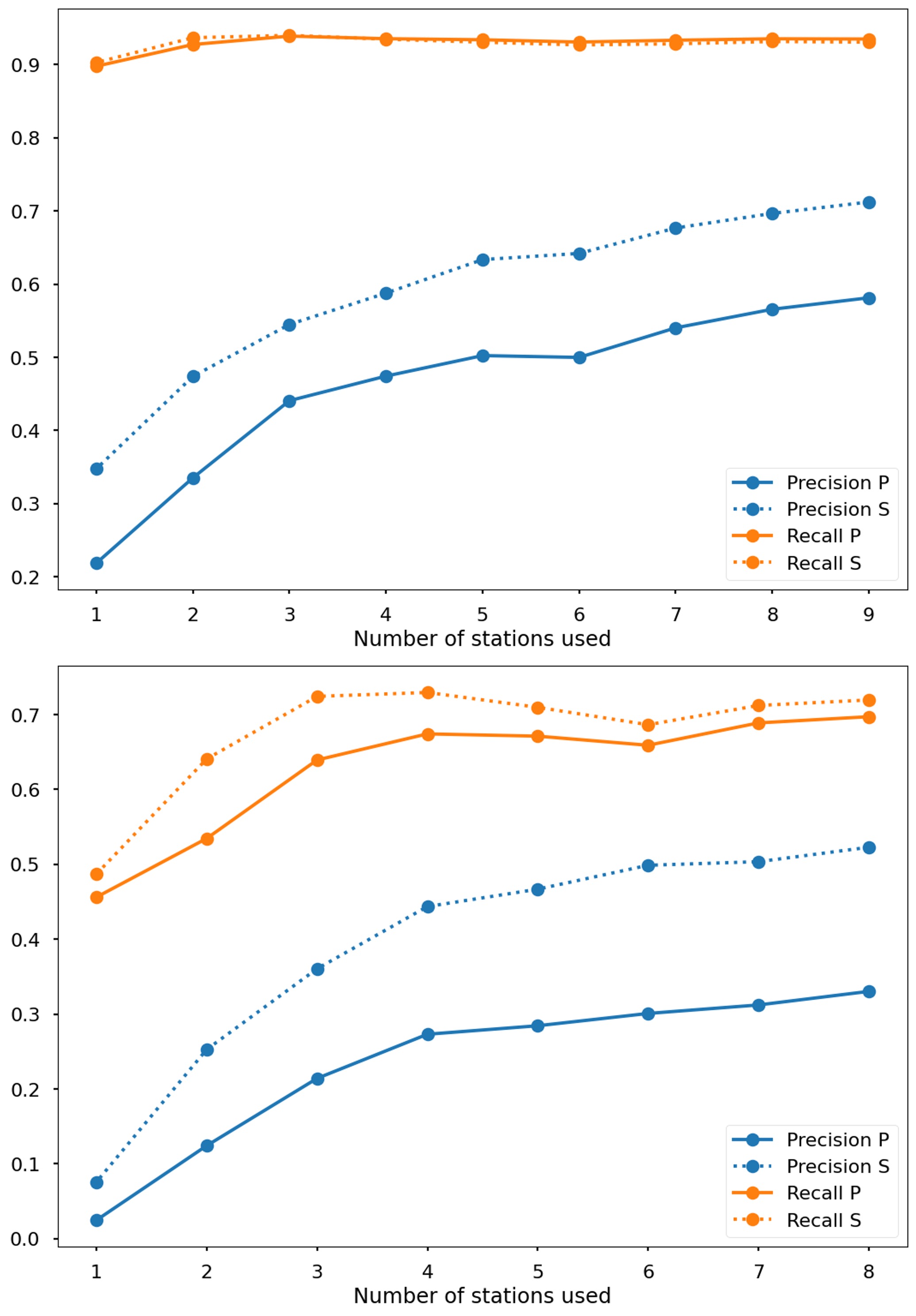}
 \caption{Adding station test with phase picking precision and recall for test 2 (top) and 4 (bottom). Blue lines are precision scores while the orange lines are recall scores. The solid lines are for P picks while the dotted lines are for S. \autoref{tab:A5} and \autoref{tab:A6} have more detailed information behind these tests. 
 }
 \label{fig:stationTest}
\end{figure}

To further understand the effects of the multi-station contributions, we conduct the experiment from the single station test and add more stations one by one for the PhaseNO model. \autoref{fig:stationTest} shows the precision and recall values of two experiments for test 2 and 4, with the first one using seismic waveforms  with a higher SNR. One common trend between the two tests is that precision are increases with the number of stations used as input. This is mainly due to PhaseNO removing many false positive localized picks if there are no picks on the other stations. This trend is very clear when increasing from 1 to 3 or 4 stations, as the precision values increase very fast. After that, adding more stations continue to increase the precision with a slower rate. Recall values from the two tests have a different pattern. Because test 2 has cleaner waveforms, even with one station as input, the recall value only decreases about 3 to 4 percent compared to using all the stations, and they are almost flat across the different number of stations as input. But for test 4, with many low SNR picks, the recall values decrease about 20\% when have only one station and increase as the number of stations increase. The recall values start to flatten out when when including more than 4 stations as input. These analyses indicate that the PhaseNO model with multi-station input can improve detection capbilities within a noisy environment. 

In this paper, we use the PhaseNO directly without any modification. As shown in many applications \citep{Chachra2022, Chai2020, Tan2018, Tang2024}, transfer learning usually helps the machine learning model to accommodate to new situations that is not included in the training dataset. We expect that using transfer learning to finetune the parameters of PhaseNO on local network setting can improve the results. We leave this as future work if we need to improve PhaseNO. 

\section*{Data Availability Statement}
The research data associated with this article are from ISTI, please contact Alex Dzubay from ISTI to request access to the data. 

\FloatBarrier

\begin{acknowledgments}
 This project RECONNECT (Real-time inducEd seismiCity fOrecasts learNiNg sEismic CaTalogs) is supported by funding bearing award number: L110-1608 from the U.S. Department of Energy’s Office of Fossil Energy and Carbon Management via the Technology Commercialization Fund (TCF), which is administered by the Office of Technology Commercialization. The TCF aims to promote the commercialization of DOE National Lab, plant, and site technologies and build out the National Lab commercialization ecosystem. This work is funded through DOE Office of Technology Transitions Core Laboratory Infrastructure for Market Readiness. Qingkai Kong’s efforts are supported both by RECONNECT project and Laboratory Directed Research \& Development at Lawrence Livermore National Laboratory, under 24-ERD-012. The work was performed under the auspices of the U.S. Department of Energy by Lawrence Livermore National Laboratory under Contract DE-AC52-07NA27344. This is LLNL Contribution Number LLNL-JRNL-2011805.
\end{acknowledgments}

\bibliographystyle{gji}
\bibliography{ref}

\begin{thebibliography}{23}
\expandafter\ifx\csname natexlab\endcsname\relax\def\natexlab#1{#1}\fi

\bibitem[Allen(1978)]{Allen1978}
Allen, R.~V., 1978.
\newblock Automatic earthquake recognition and timing from single traces, {\it Bulletin of the seismological society of America\/}, {\bf 68}(5), 1521--1532.

\bibitem[Chachra et~al.(2022)Chachra, Kong, Huang, Korlakunta, Grannen, Robson, \& Allen]{Chachra2022}
Chachra, G., Kong, Q., Huang, J., Korlakunta, S., Grannen, J., Robson, A., \& Allen, R.~M., 2022.
\newblock Detecting damaged buildings using real-time crowdsourced images and transfer learning, {\it Scientific Reports\/}, {\bf 12}(1), 8968.

\bibitem[Chai et~al.(2020)Chai, Maceira, Santos-Villalobos, Venkatakrishnan, Schoenball, Zhu, Beroza, Thurber, \& Team]{Chai2020}
Chai, C., Maceira, M., Santos-Villalobos, H.~J., Venkatakrishnan, S.~V., Schoenball, M., Zhu, W., Beroza, G.~C., Thurber, C., \& Team, E. G. S.~C., 2020.
\newblock Using a deep neural network and transfer learning to bridge scales for seismic phase picking, {\it Geophysical Research Letters\/}, {\bf 47}(16), e2020GL088651.

\bibitem[Dokht et~al.(2019)Dokht, Kao, Visser, \& Smith]{Dokht2019}
Dokht, R.~M., Kao, H., Visser, R., \& Smith, B., 2019.
\newblock Seismic event and phase detection using time–frequency representation and convolutional neural networks, {\it Seismological Research Letters\/}, {\bf 90}(2A), 481--490.

\bibitem[Feng et~al.(2022)Feng, Mohanna, \& Meng]{Feng2022}
Feng, T., Mohanna, S., \& Meng, L., 2022.
\newblock Edgephase: A deep learning model for multi‐station seismic phase picking, {\it Geochemistry, Geophysics, Geosystems\/}, {\bf 23}(11), e2022GC010453.

\bibitem[Johnson \& Johnson(2022)]{Johnson2022}
Johnson, C.~W. \& Johnson, P.~A., 2022.
\newblock Eqdetect: Earthquake phase arrivals and first motion polarity applying deep learning, {\it ESS Open Archive eprints\/}, {\bf 105}, essoar. 10511191.

\bibitem[Kovachki et~al.(2021)Kovachki, Li, Liu, Azizzadenesheli, Bhattacharya, Stuart, \& Anandkumar]{Kovachki2021}
Kovachki, N., Li, Z., Liu, B., Azizzadenesheli, K., Bhattacharya, K., Stuart, A., \& Anandkumar, A., 2021.
\newblock Neural operator: Learning maps between function spaces.

\bibitem[Li et~al.(2020{\natexlab{a}})Li, Kovachki, Azizzadenesheli, Liu, Bhattacharya, Stuart, \& Anandkumar]{Li2020a}
Li, Z., Kovachki, N., Azizzadenesheli, K., Liu, B., Bhattacharya, K., Stuart, A., \& Anandkumar, A., 2020{\natexlab{a}}.
\newblock Fourier neural operator for parametric partial differential equations.

\bibitem[Li et~al.(2020{\natexlab{b}})Li, Kovachki, Azizzadenesheli, Liu, Bhattacharya, Stuart, \& Anandkumar]{Li2020b}
Li, Z., Kovachki, N., Azizzadenesheli, K., Liu, B., Bhattacharya, K., Stuart, A., \& Anandkumar, A., 2020{\natexlab{b}}.
\newblock Neural operator: Graph kernel network for partial differential equations, {\it arXiv.org\/}.

\bibitem[Mousavi et~al.(2020)Mousavi, Ellsworth, Zhu, Chuang, \& Beroza]{Mousavi2020}
Mousavi, S.~M., Ellsworth, W.~L., Zhu, W., Chuang, L.~Y., \& Beroza, G.~C., 2020.
\newblock Earthquake transformer—an attentive deep-learning model for simultaneous earthquake detection and phase picking, {\it Nature communications\/}, {\bf 11}(1), 3952.

\bibitem[Münchmeyer(2023)]{Munchmeyer2023}
Münchmeyer, J., 2023.
\newblock Pyocto: A high-throughput seismic phase associator, {\it arXiv preprint arXiv:2310.11157\/}.

\bibitem[Ronneberger et~al.(2015)Ronneberger, Fischer, \& Brox]{Ronneberger2015}
Ronneberger, O., Fischer, P., \& Brox, T., 2015.
\newblock U-net: Convolutional networks for biomedical image segmentation, in {\em International Conference on Medical image computing and computer-assisted intervention\/}, pp. 234--241, Springer.

\bibitem[Ross et~al.(2018)Ross, Meier, Hauksson, \& Heaton]{Ross2018}
Ross, Z.~E., Meier, M., Hauksson, E., \& Heaton, T.~H., 2018.
\newblock Generalized seismic phase detection with deep learning, {\it Bulletin of the Seismological Society of America\/}, {\bf 108}(5A), 2894--2901.

\bibitem[Sun et~al.(2023)Sun, Ross, Zhu, \& Azizzadenesheli]{Sun2023}
Sun, H., Ross, Z.~E., Zhu, W., \& Azizzadenesheli, K., 2023.
\newblock Phase neural operator for multi‐station picking of seismic arrivals, {\it Geophysical Research Letters\/}, {\bf 50}(24), e2023GL106434.

\bibitem[Takanami \& Kitagawa(1988)]{Takanami1988}
Takanami, T. \& Kitagawa, G., 1988.
\newblock A new efficient procedure for the estimation of onset times of seismic waves, {\it Journal of Physics of the Earth\/}, {\bf 36}(6), 267--290.

\bibitem[Tan et~al.(2018)Tan, Sun, Kong, Zhang, Yang, \& Liu]{Tan2018}
Tan, C., Sun, F., Kong, T., Zhang, W., Yang, C., \& Liu, C., 2018.
\newblock A survey on deep transfer learning, Lecture Notes in Computer Science, pp. 270--279, Springer International Publishing, Cham.

\bibitem[Tang et~al.(2024)Tang, Kong, \& Morris]{Tang2024}
Tang, H., Kong, Q., \& Morris, J.~P., 2024.
\newblock Multi-fidelity fourier neural operator for fast modeling of large-scale geological carbon storage, {\it Journal of Hydrology\/}, {\bf 629}, 130641.

\bibitem[Woollam et~al.(2022)Woollam, Munchmeyer, Tilmann, Rietbrock, Lange, Bornstein, Diehl, Giunchi, Haslinger, Jozinoviƒá, Michelini, Saul, \& Soto]{Woollam2022}
Woollam, J., Munchmeyer, J., Tilmann, F., Rietbrock, A., Lange, D., Bornstein, T., Diehl, T., Giunchi, C., Haslinger, F., Jozinoviƒá, D., Michelini, A., Saul, J., \& Soto, H., 2022.
\newblock Seisbench — a toolbox for machine learning in seismology, {\it Seismological Research Letters\/}, {\bf 93}(3), 1695--1709.

\bibitem[Zhang et~al.(2019)Zhang, Ellsworth, \& Beroza]{Zhang2019}
Zhang, M., Ellsworth, W.~L., \& Beroza, G.~C., 2019.
\newblock Rapid earthquake association and location, {\it Seismological Research Letters\/}, {\bf 90}(6), 2276--2284.

\bibitem[Zhou et~al.(2019)Zhou, Yue, Kong, \& Zhou]{Zhou2019}
Zhou, Y., Yue, H., Kong, Q., \& Zhou, S., 2019.
\newblock Hybrid event detection and phase picking algorithm using convolutional and recurrent neural networks, {\it Seismological Research Letters\/}, {\bf 90}(3), 1079--1087.

\bibitem[Zhu \& Beroza(2019)]{Zhu2019}
Zhu, W. \& Beroza, G.~C., 2019.
\newblock Phasenet: a deep-neural-network-based seismic arrival-time picking method, {\it Geophysical Journal International\/}, {\bf 216}(1), 261--273.

\bibitem[Zhu et~al.(2022{\natexlab{a}})Zhu, McBrearty, Mousavi, Ellsworth, \& Beroza]{Zhu2022a}
Zhu, W., McBrearty, I.~W., Mousavi, S.~M., Ellsworth, W.~L., \& Beroza, G.~C., 2022{\natexlab{a}}.
\newblock Earthquake phase association using a bayesian gaussian mixture model, {\it Journal of Geophysical Research: Solid Earth\/}, {\bf 127}(5), e2021JB023249.

\bibitem[Zhu et~al.(2022{\natexlab{b}})Zhu, Tai, Mousavi, Bailis, \& Beroza]{Zhu2022b}
Zhu, W., Tai, K.~S., Mousavi, S.~M., Bailis, P., \& Beroza, G.~C., 2022{\natexlab{b}}.
\newblock An end‐to‐end earthquake detection method for joint phase picking and association using deep learning, {\it Journal of Geophysical Research: Solid Earth\/}, {\bf 127}(3), e2021JB023283.

\end{thebibliography}

\FloatBarrier
\appendix

\section{Supplementary Figures and Tables}\label{appendix}

\subsection{Supplementary figures}

\begin{figure}
 \centering
 \includegraphics[width=.5\textwidth]{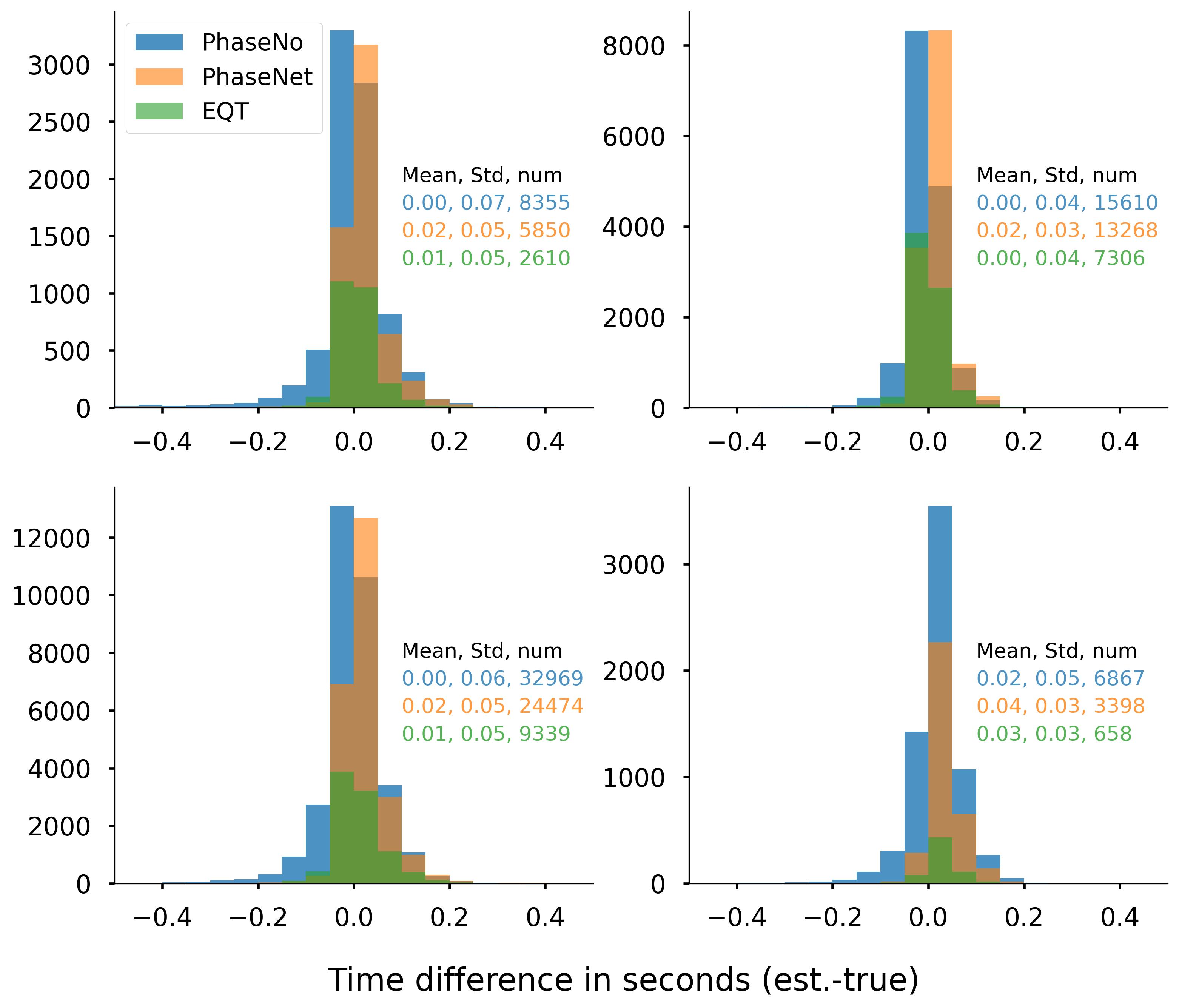}
 \caption{S wave picking time differences comparing to the manual pickings.
 }
 \label{fig:A1}
\end{figure}

\begin{figure}
 \centering
 \includegraphics[width=.5\textwidth]{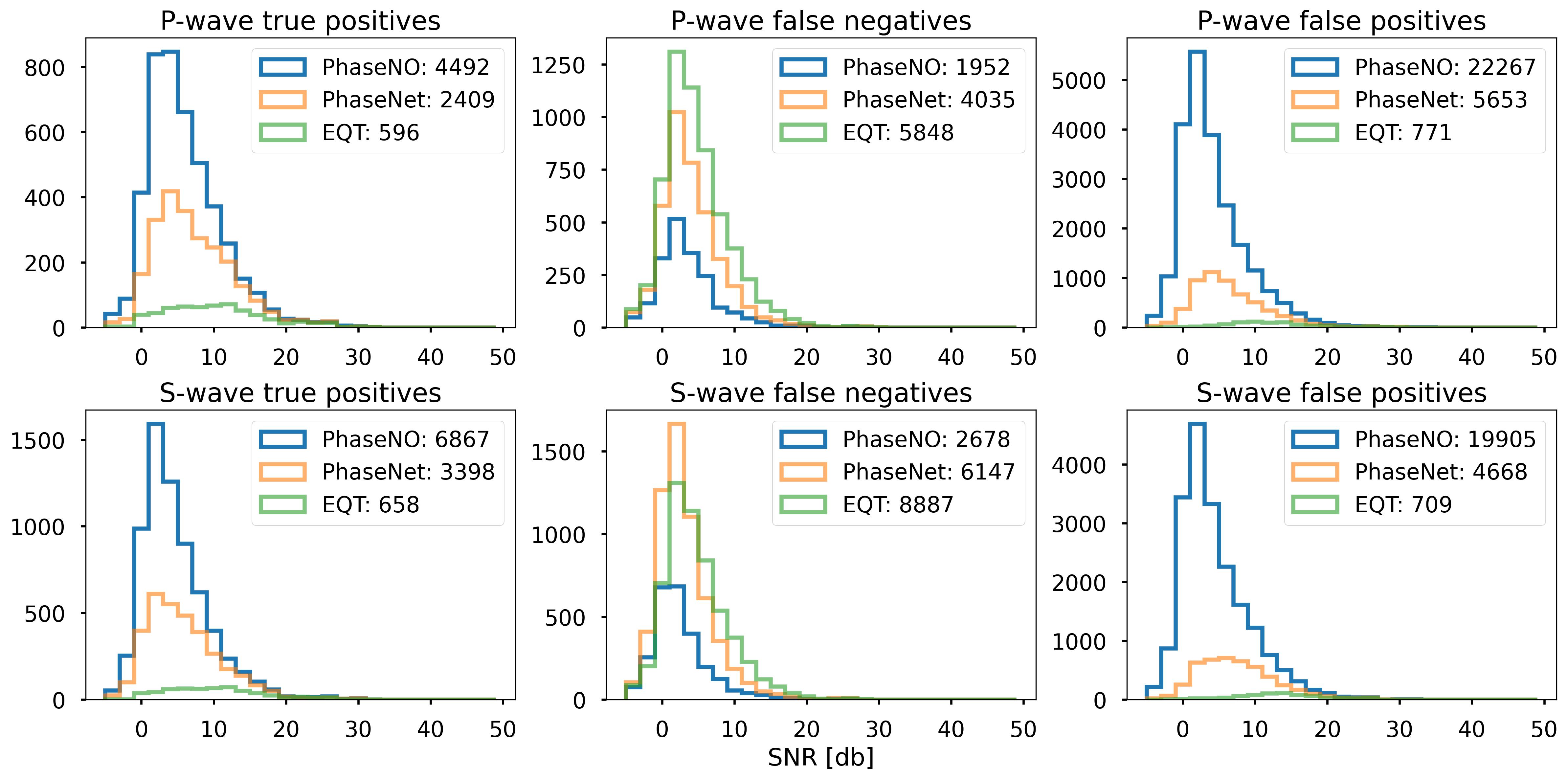}
 \caption{Test 2 phase true positives, false negatives, and false positives distributions for both P and S phases against signal noise ratio.
 }
 \label{fig:A2}
\end{figure}

\begin{figure}
 \centering
 \includegraphics[width=.5\textwidth]{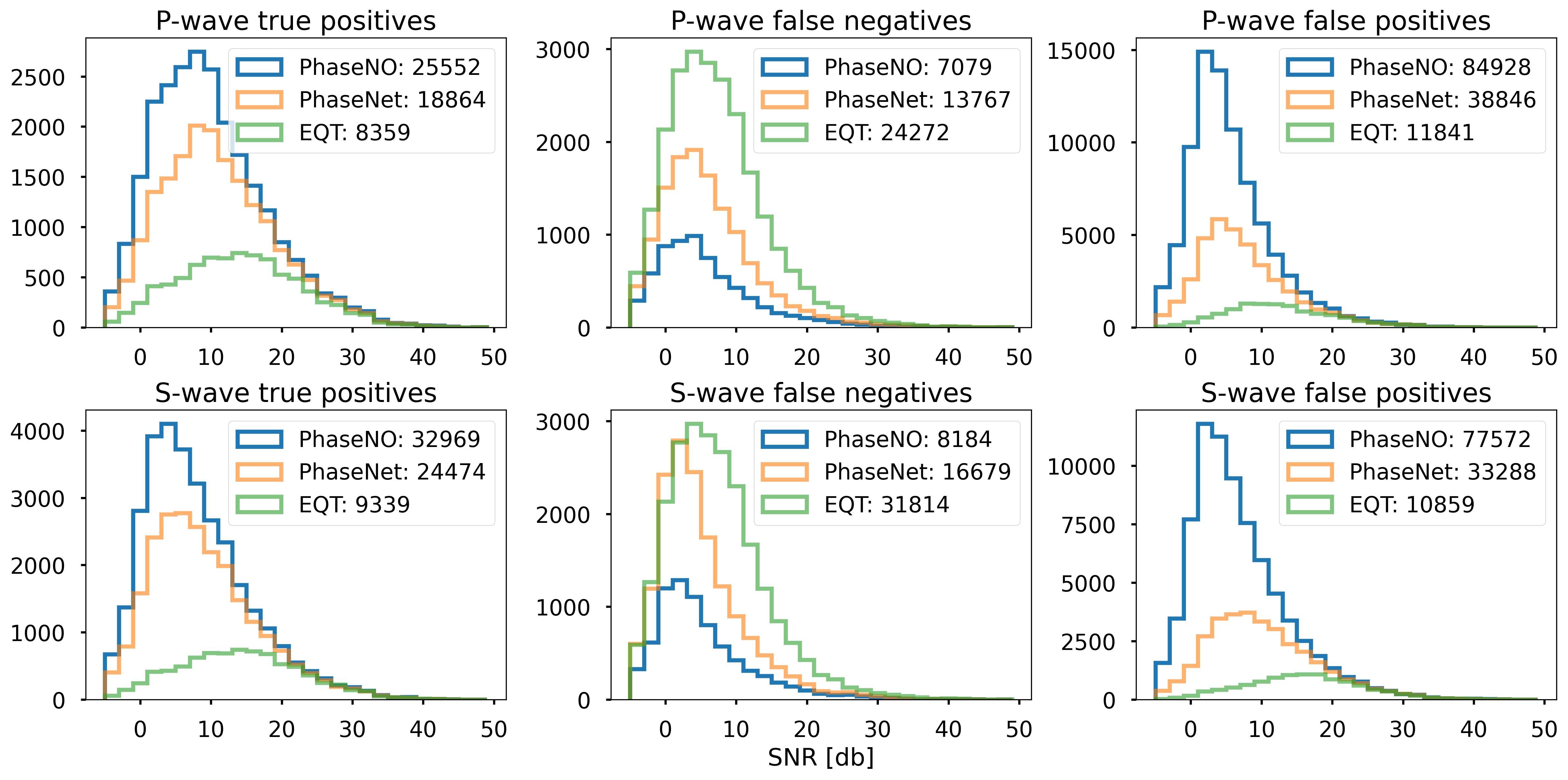}
 \caption{Test 3 phase true positives, false negatives, and false positives distributions for both P and S phases against signal noise ratio.
 }
 \label{fig:A3}
\end{figure}

\begin{figure}
 \centering
 \includegraphics[width=.5\textwidth]{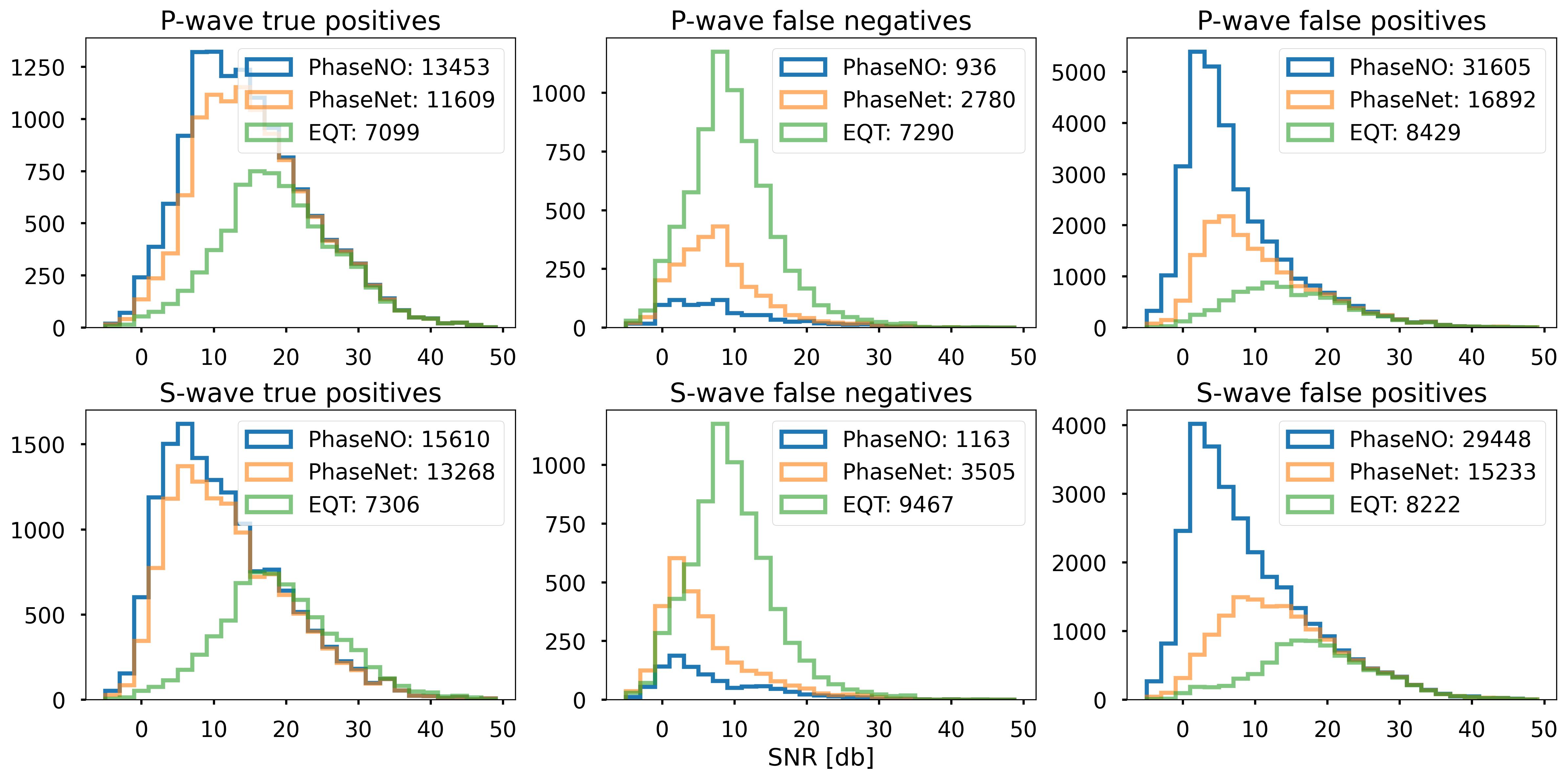}
 \caption{Test 4 phase true positives, false negatives, and false positives distributions for both P and S phases against signal noise ratio.
 }
 \label{fig:A4}
\end{figure}

\begin{figure}
 \centering
 \includegraphics[width=.5\textwidth]{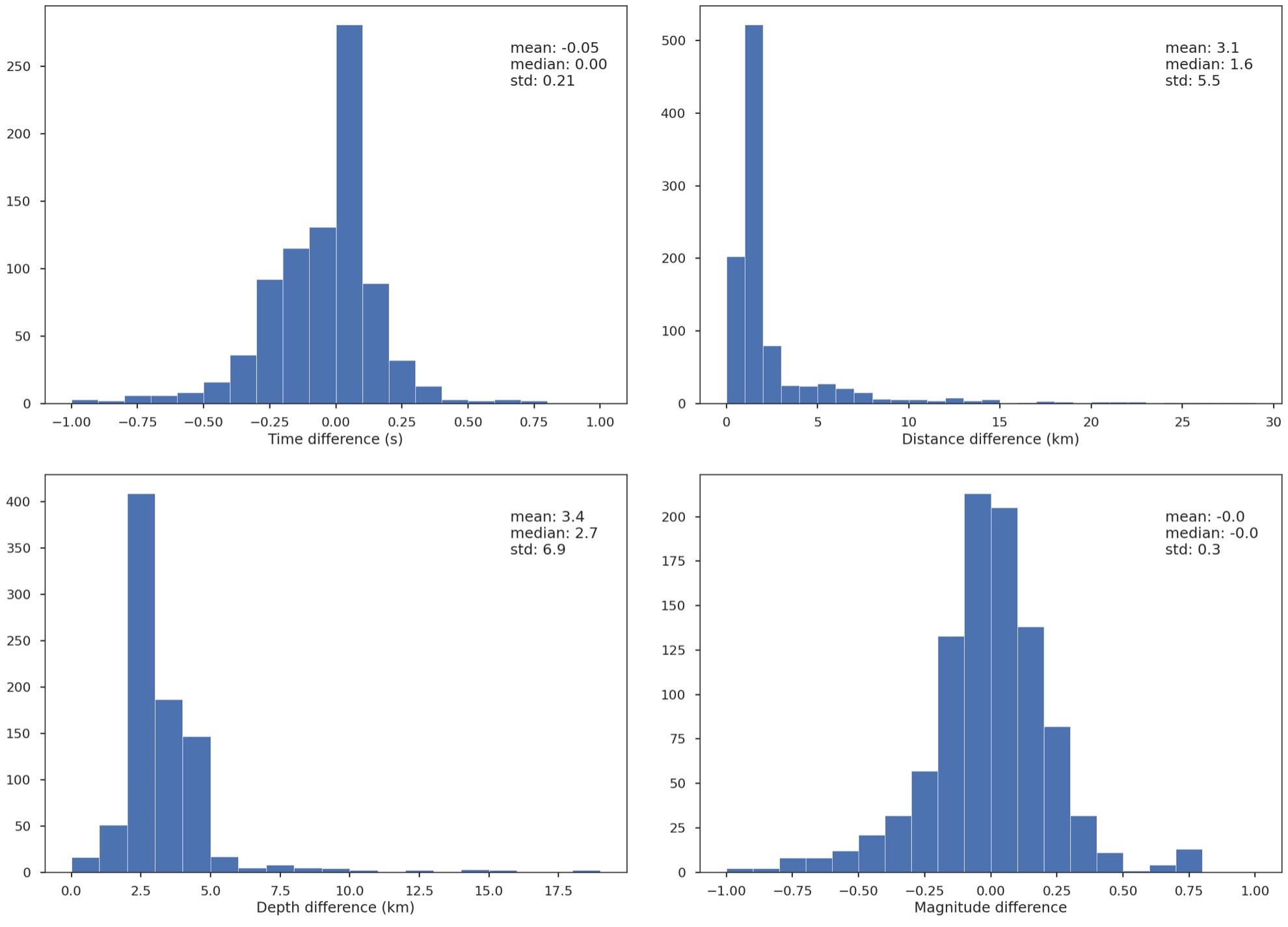}
 \caption{Test 1 matched events differences for origin time, epicentral distance, depth, and magnitude against manual catalog.
 }
 \label{fig:A5}
\end{figure}

\begin{figure}
 \centering
 \includegraphics[width=.5\textwidth]{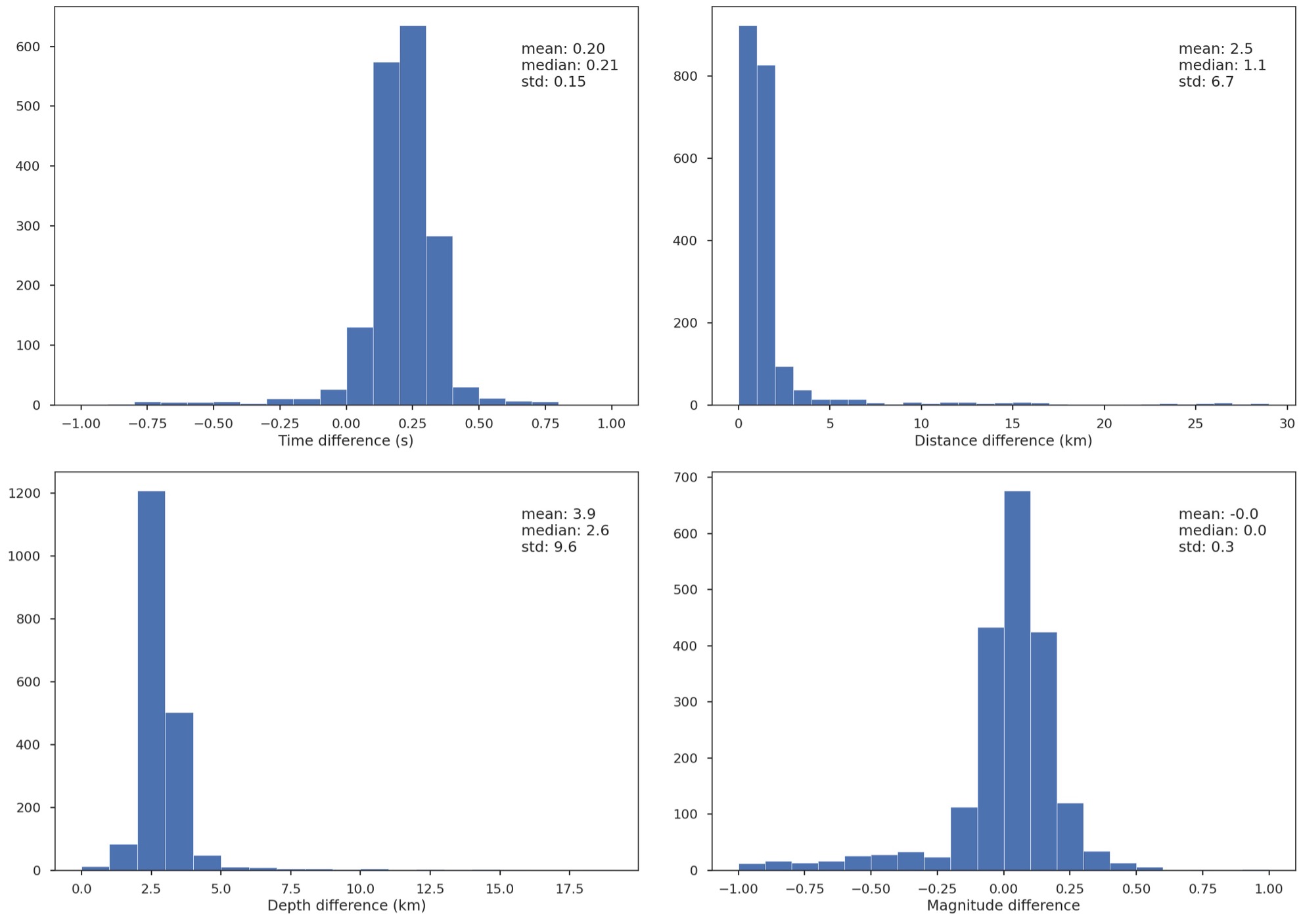}
 \caption{Test 2 matched events differences for origin time, epicentral distance, depth, and magnitude against manual catalog.
 }
 \label{fig:A6}
\end{figure}

\begin{figure}
 \centering
 \includegraphics[width=.5\textwidth]{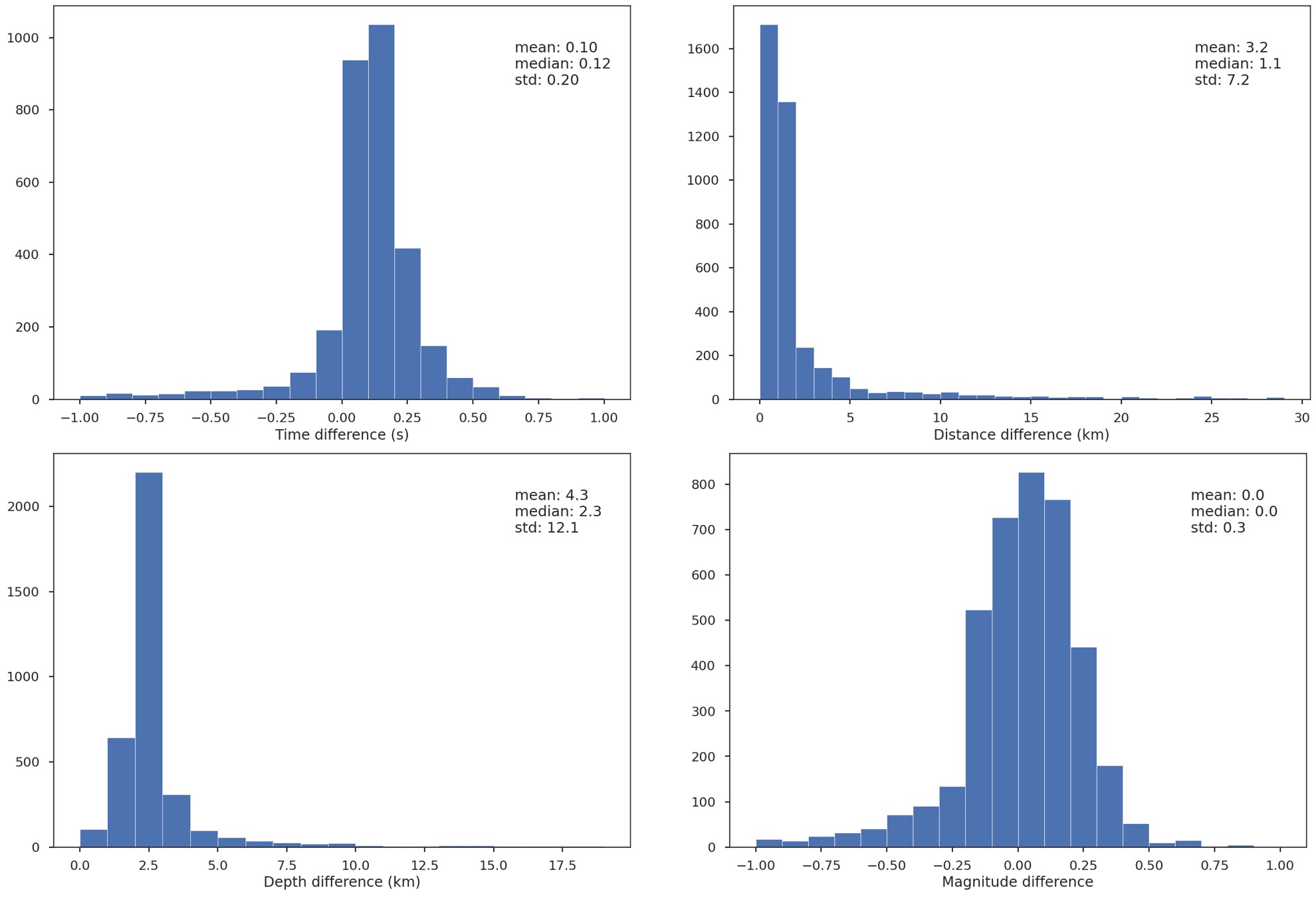}
 \caption{Test 3 matched events differences for origin time, epicentral distance, depth, and magnitude against manual catalog.
 }
 \label{fig:A7}
\end{figure}

\begin{figure}
 \centering
 \includegraphics[width=.5\textwidth]{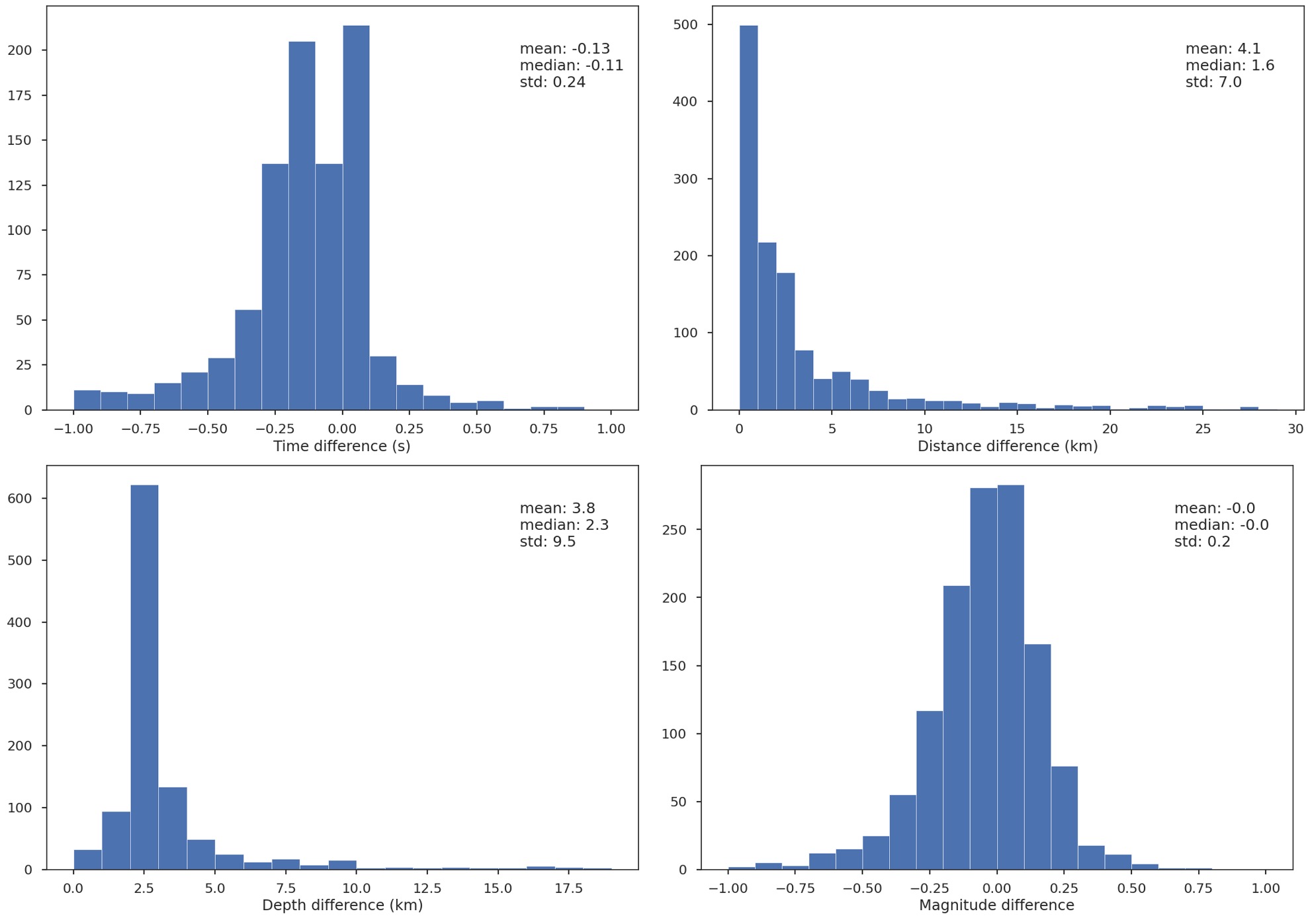}
 \caption{Test 4 matched events differences for origin time, epicentral distance, depth, and magnitude against manual catalog.
 }
 \label{fig:A8}
\end{figure}

\begin{figure}
 \centering
 \includegraphics[width=.5\textwidth]{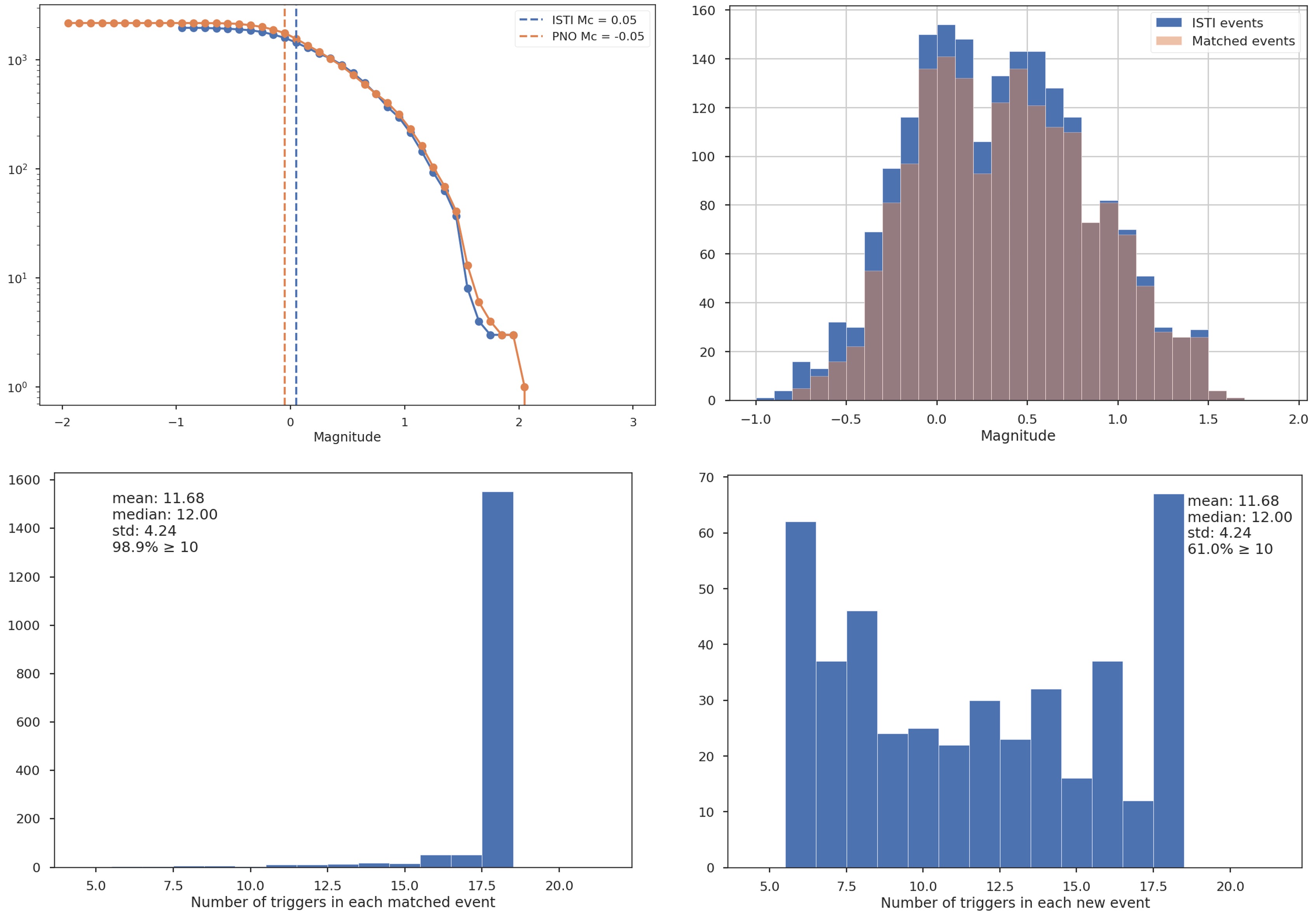}
 \caption{More metrics for the associated events for test 2. Top left panel shows the magnitude GR relationship and magnitude completeness for both PNO and ISTI catalogs. The top right panel shows the magnitude distribution for ISTI catalog as well as for the matched events from PNO. The bottom two panels show the trigger distributions for each matched events and the new detected events.
 }
 \label{fig:A9}
\end{figure}

\begin{figure}
 \centering
 \includegraphics[width=.5\textwidth]{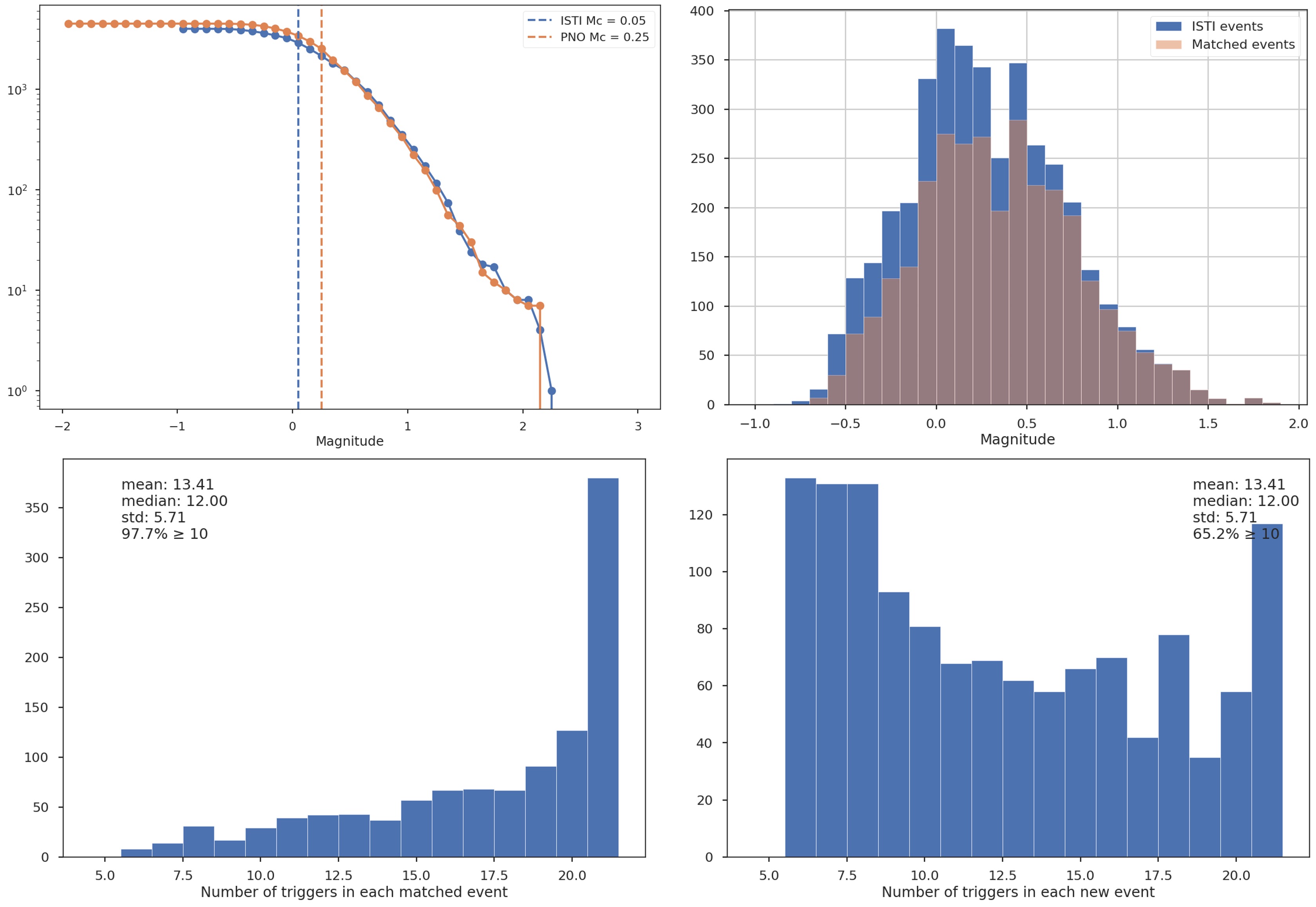}
 \caption{More metrics for the associated events for test 3.
 }
 \label{fig:A10}
\end{figure}

\begin{figure}
 \centering
 \includegraphics[width=.5\textwidth]{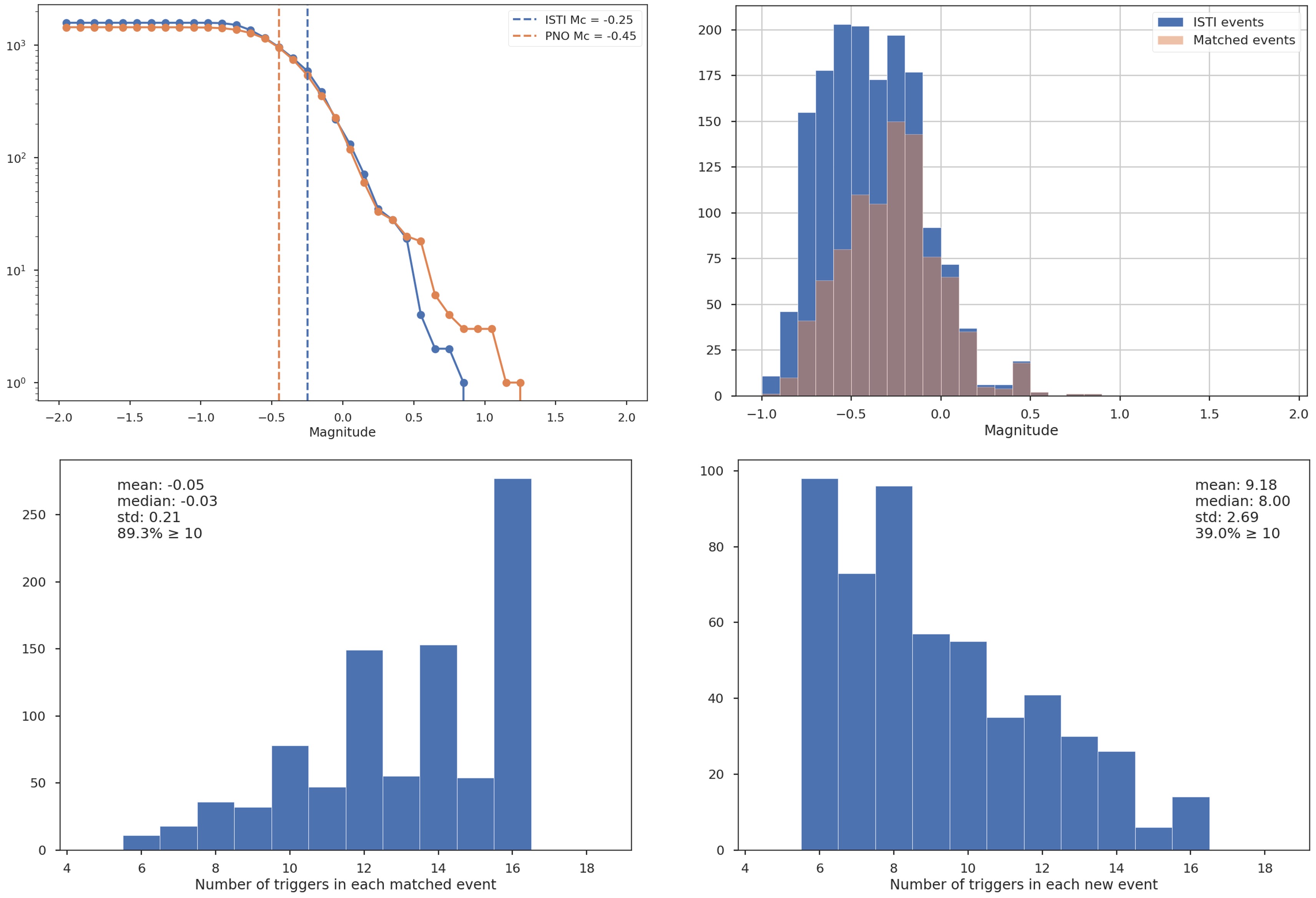}
 \caption{More metrics for the associated events for test 4.
 }
 \label{fig:A11}
\end{figure}

\FloatBarrier

\subsection{Supplementary tables}
\begin{table*}
    \centering
    
    \begin{tabular}{|c|c|c|c|c|c|}
        \hline
        \textbf{Test} & \textbf{Name} & \textbf{ISTI} & \textbf{PhaseNO} & \textbf{PhaseNet} & \textbf{EQT} \\
        \hline
        \multirow{6}{*}{1} & Total & 19153 & 26026 & 12982 & 5560 \\
        & P/S & 8296/10857 & 13437/12589 & 6228/6754 & 2664/2896 \\
        & Matched P/S & / & 6462/8355 & 4601/5850 & 2345/2610 \\
        & Precision (P/S) & / & 0.48/0.66 & 0.74/0.87 & 0.88/0.90 \\
        & Recall (P/S) & / & 0.78/0.77 & 0.55/0.54 & 0.28/0.24 \\
        & f1 (P/S) & / & 0.59/0.71 & 0.63/0.66 & 0.43/0.38 \\
        \hline
        \multirow{6}{*}{2} & Total & 31162 & 45058 & 28501 & 15528 \\
        & P/S & 14389/16773 & 23141/21917 & 13959/14542 & 7555/7973 \\
        & Matched P/S & / & 13453/15610 & 11609/13268 & 7099/7306 \\
        & Precision (P/S) & / & 0.58/0.71 & 0.83/0.91 & 0.94/0.92 \\
        & Recall (P/S) & / & 0.93/0.93 & 0.81/0.79 & 0.49/0.44 \\
        & f1 (P/S) & / & 0.72/0.81 & 0.82/0.85 & 0.65/0.59 \\
        \hline
        \multirow{6}{*}{3} & Total & 73784 & 110405 & 57672 & 20198 \\
        & P/S & 32631/41153 & 56626/53779 & 27012/30660 & 9408/10790 \\
        & Matched P & / & 25552/32969 & 18864/24474 & 8359/9339 \\
        & Precision (P/S) & / & 0.45/0.61 & 0.70/0.80 & 0.89/0.87 \\
        & Recall (P/S) & / & 0.78/0.80 & 0.58/0.59 & 0.26/0.23 \\
        & f1 (P/S) & / & 0.57/0.69 & 0.63/0.68 & 0.40/0.36 \\
        \hline
        \multirow{6}{*}{4} & Total & 16005 & 26804 & 8057 & 1350 \\
        & P/S & 6453/9552 & 13623/13181 & 3671/4386 & 625/725 \\
        & Matched P & / & 4492/6867 & 2409/3398 & 596/658 \\
        & Precision (P/S) & / & 0.33/0.52 & 0.65/0.78 & 0.93/0.90 \\
        & Recall (P/S) & / & 0.70/0.72 & 0.37/0.36 & 0.09/0.07 \\
        & f1 (P/S) & / & 0.45/0.61 & 0.48/0.49 & 0.17/0.13 \\
        \hline

    \end{tabular}
    \caption{Matched phase time residual comparing to manually picked phases}
    \label{tab:A1}
\end{table*}

\begin{table*}
    \centering
    
    \begin{tabular}{|c|l|c|c|c|c|c|c|}
        \hline
        \textbf{Test} & \textbf{Model} & \textbf{mean\_p} & \textbf{std\_p} & \textbf{Num\_p} & \textbf{Mean\_s} & \textbf{Std\_s} & \textbf{Num\_s} \\
        \hline
        1 & PhaseNO   & 0.00 & 0.08 & 6462 & 0.00 & 0.07 & 8355 \\
          & PhaseNet  & 0.02 & 0.05 & 4601 & 0.02 & 0.05 & 5850 \\
          & EQT       & 0.01 & 0.05 & 2345 & 0.01 & 0.05 & 2610 \\
        \hline
        2 & PhaseNO   & 0.00 & 0.05 & 13453 & 0.00 & 0.04 & 15610 \\
          & PhaseNet  & 0.01 & 0.04 & 11609 & 0.02 & 0.03 & 13268 \\
          & EQT       & 0.01 & 0.04 & 7099  & 0.00 & 0.04 & 7306 \\
        \hline
        3 & PhaseNO   & 0.00 & 0.08 & 25552 & 0.00 & 0.06 & 32969 \\
          & PhaseNet  & 0.02 & 0.04 & 18864 & 0.02 & 0.05 & 24474 \\
          & EQT       & 0.01 & 0.05 & 8359  & 0.01 & 0.05 & 9339 \\
        \hline
        4 & PhaseNO   & 0.01 & 0.09 & 4492  & 0.02 & 0.05 & 6867 \\
          & PhaseNet  & 0.04 & 0.06 & 2409  & 0.04 & 0.04 & 3398 \\
          & EQT       & 0.03 & 0.05 & 596   & 0.03 & 0.04 & 658 \\
        \hline
        \label{tab:A2}
    \end{tabular}
    \caption{Matched phase time residual comparing to manually picked phases}
\end{table*}

\begin{table*}
    \centering
    \begin{tabular}{|c|c|c|c|c|c|c|c|c|c|c|c|c|}
        \hline
        \textbf{Test} & \textbf{model} & \textbf{\# detection} & \textbf{\# matched} & \textbf{\# missing} & \textbf{\# new detection} & \textbf{Associate (percent)} & \textbf{Total phases} & \textbf{Precision} & \textbf{Recall} & \textbf{f1} \\
        \hline
        \multirow{3}{*}{1} & ISTI & 1070 & / & / & / & / & 19153 & & & \\
         & pn & 629 & 604 & 466 & 25 & 10482 (79\%) & 13255 & 0.960 & 0.564 & 0.711 \\
         & pno & 1041 & 840 & 230 & 201 & 19677 (76\%) & 26026 & 0.807 & 0.785 & 0.796 \\
        \hline
        \multirow{3}{*}{2} & ISTI & 1966 & / & / & / & / & 31162 & & & \\
         & pn & 1718 & 1640 & 326 & 78 & 26002 (86\%) & 30077 & 0.955 & 0.834 & 0.890 \\
         & pno & 2177 & 1744 & 222 & 433 & 35749 (79\%) & 45058 & 0.801 & 0.887 & 0.842 \\
        \hline
        \multirow{3}{*}{3} & ISTI & 3991 & / & / & / & / & 73784 & & & \\
         & pn & 2885 & 2511 & 1480 & 374 & 46210 (79\%) & 58320 & 0.870 & 0.629 & 0.730 \\
         & pno & 4494 & 3090 & 901 & 1404 & 85651 (78\%) & 110405 & 0.688 & 0.774 & 0.728 \\
        \hline
        \multirow{3}{*}{4} & ISTI & 1579 & / & / & / & / & 16005 & & & \\
         & pn & 464 & 419 & 1160 & 45 & 4524 (55\%) & 8278 & 0.903 & 0.265 & 0.410 \\
         & pno & 1441 & 910 & 669 & 531 & 16827 (63\%) & 26740 & 0.632 & 0.576 & 0.603 \\
        \hline
    \end{tabular}
    \caption{Associated events metrics with matching threshold 1s.}
    \label{tab:A3}
\end{table*}

\begin{table*}
    \centering
    \begin{tabular}{|c|c|c|c|c|}
        \hline
        \textbf{Test} & \textbf{Name} & \textbf{ISTI} & \textbf{PhaseNO} & \textbf{PhaseNO1} \\
        \hline
        \multirow{6}{*}{1}
        & Total & 19153 & 26026 & 68866 \\
        & P/S & 8296/10857 & 13437/12589 & 37830/31036 \\
        & Matched P/S & / & 6462/8355 & 5750/7725 \\
        & Precision (P/S) & / & 0.48/0.66 & 0.15/0.25 \\
        & Recall (P/S) & / & 0.78/0.77 & 0.69/0.71 \\
        & f1 (P/S) & / & 0.59/0.71 & 0.25/0.37 \\
        \hline
        \multirow{6}{*}{2}
        & Total & 31162 & 45058 & 99418 \\
        & P/S & 14389/16773 & 23141/21917 & 55162/44256 \\
        & Matched P/S & / & 13453/15610 & 12707/15119 \\
        & Precision (P/S) & / & 0.58/0.71 & 0.23/0.34 \\
        & Recall (P/S) & / & 0.93/0.93 & 0.88/0.90 \\
        & f1 (P/S) & / & 0.72/0.81 & 0.37/0.50 \\
        \hline
        \multirow{6}{*}{3}
        & Total & 73784 & 110405 & 235311 \\
        & P/S & 32631/41153 & 56626/53779 & 128151/107160 \\
        & Matched P & / & 25552/32969 & 21398/29138 \\
        & Precision (P/S) & / & 0.45/0.61 & 0.17/0.27 \\
        & Recall (P/S) & / & 0.78/0.80 & 0.66/0.71 \\
        & f1 (P/S) & / & 0.57/0.69 & 0.27/0.39 \\
        \hline
        \multirow{6}{*}{4}
        & Total & 16005 & 26804 & 62633 \\
        & P & 6453/9552 & 13623/13181 & 34282/28351 \\
        & Matched P & / & 4492/6867 & 3455/6451 \\
        & Precision (P/S) & / & 0.33/0.52 & 0.10/0.23 \\
        & Recall (P/S) & / & 0.70/0.72 & 0.54/0.68 \\
        & f1 (P/S) & / & 0.45/0.61 & 0.17/0.34 \\
        \hline
    \end{tabular}
    \caption{Phase picking metrics for PhaseNO and PhaseNO1}
    \label{tab:A4}
\end{table*}

\begin{table*}
    \centering
    \begin{tabular}{|c|c|c|c|c|c|}

        \hline
        \# of Stations & 9 & 8 & 7 & 6 & 5 \\
        \hline
        Stations & \shortstack{S01, S09, S04, S08, \\ S02, S05, S03, S06, S07} & \shortstack{S01, S09, S08, S02, \\ S05, S03, S06, S07} & \shortstack{S09, S08, S02, S05, \\ S06, S07} & \shortstack{S09, S02, S05, S03, \\ S07} & \shortstack{S02, S05, S03, S06, \\ S07} \\
        \hline
        ISTI P/S & 14389/16773 & 12884/14967 & 11295/13097 & 9566/11194 & 7932/9267 \\
        \hline
        PNO P/S & 23141/21917 & 21303/20013 & 19550/18197 & 17920/16457 & 14746/13612 \\
        \hline
        Matched P/S & 13453/15610 & 12050/13942 & 10575/12199 & 8922/10408 & 7408/9267 \\
        \hline
        Precision (P/S) & 0.58/0.71 & 0.57/0.70 & 0.54/0.67 & 0.50/0.63 & 0.50/0.63 \\
        \hline
        Recall (P/S) & 0.93/0.93 & 0.94/0.93 & 0.94/0.93 & 0.93/0.93 & 0.93/0.93 \\
        \hline
        f1 (P/S) & 0.72/0.81 & 0.70/0.80 & 0.69/0.78 & 0.65/0.75 & 0.65/0.75 \\
        \hline
        \# of Stations & 4 & 3 & 2 & 1 & \\
        \hline
        Stations & S05, S03, S06, S07 & S03, S06, S07 & S06, S07 & S07 & \\
        \hline
        ISTI P/S & 6515/7459 & 5028/5713 & 3114/3844 & 9566/11194 & \\
        \hline
        PNO P/S & 12847/11862 & 10710/9854 & 8611/7598 & 17920/16457 & \\
        \hline
        Matched P/S & 6094/6972 & 4721/5370 & 2888/3601 & 8922/10408 & \\
        \hline
        Precision (P/S) & 0.47/0.59 & 0.44/0.54 & 0.34/0.47 & 0.22/0.35 & \\
        \hline
        Recall (P/S) & 0.94/0.93 & 0.94/0.94 & 0.93/0.94 & 0.90/0.90 & \\
        \hline
        f1 (P/S) & 0.63/0.72 & 0.60/0.69 & 0.49/0.63 & 0.35/0.50 & \\
        \hline
    \end{tabular}
    \caption{Station adding tests for test 2}
    \label{tab:A5}
\end{table*}

\begin{table*}
    \centering
    \begin{tabular}{|c|c|c|c|c|}
        \hline
        \# of Stations & 8 & 7 & 6 & 5 \\
        \hline
        Stations & S01, S04, S08, S02, S05, S03, S06, S07 & S01, S08, S02, S05, S03, S06, S07 & S08, S02, S05, S03, S06, S07 & S02, S05, S03, S06, S07 \\
        \hline
        ISTI P/S & 6453/9552 & 5501/8069 & 4668/6817 & 4357/6168 \\
        \hline
        PNO P/S & 13623/13181 & 12140/11413 & 10227/9383 & 10283/9378 \\
        \hline
        Matched P/S & 4492/6867 & 3789/5746 & 3076/4679 & 2924/4377 \\
        \hline
        Precision (P/S) & 0.33/0.52 & 0.31/0.50 & 0.30/0.50 & 0.28/0.47 \\
        \hline
        Recall (P/S) & 0.70/0.72 & 0.69/0.71 & 0.66/0.69 & 0.67/0.71 \\
        \hline
        f1 (P/S) & 0.45/0.61 & 0.43/0.59 & 0.41/0.58 & 0.40/0.56 \\
        \hline
        \# of Stations & 4 & 3 & 2 & 1 \\
        \hline
        Stations & S05, S03, S06, S07 & S03, S06, S07 & S06, S07 & S07 \\
        \hline
        ISTI P/S & 3741/5160 & 2739/3626 & 1463/2057 & 239/536 \\
        \hline
        PNO P/S & 9238/8486 & 8181/7285 & 6295/5216 & 4425/3450 \\
        \hline
        Matched P/S & 2522/3764 & 1751/2626 & 782/1318 & 109/261 \\
        \hline
        Precision (P/S) & 0.27/0.44 & 0.21/0.36 & 0.12/0.25 & 0.02/0.08 \\
        \hline
        Recall (P/S) & 0.67/0.73 & 0.64/0.72 & 0.53/0.64 & 0.46/0.49 \\
        \hline
        f1 (P/S) & 0.39/0.55 & 0.32/0.48 & 0.20/0.36 & 0.05/0.13 \\
        \hline
    \end{tabular}
    \caption{Station adding tests for test 4}
    \label{tab:A6}
\end{table*}

\end{document}